\documentclass[journal ]{new-aiaa}
\usepackage[utf8]{inputenc}
\usepackage{textcomp}

\usepackage{graphicx}
\usepackage{amsmath}
\usepackage[version=4]{mhchem}
\usepackage{siunitx}
\usepackage{longtable,tabularx}
\title{Acoustic Loading Beneath High-Speed Flow Over a Compression Ramp at Different Angles}

\author{Ritu Raj Kumar \footnote{PhD Scholar, Department of Aerospace Engineering, IIT Madras, ae22d016@smail.iitm.ac.in.} and Nagabhushana Rao Vadlamani.\footnote{Associate Professor, Department of Aerospace Engineering, IIT Madras, nrv@iitm.ac.in, Senior Member AIAA.}}
\affil{Department of Aerospace Engineering, Indian Institute of Technology Madras, Chennai, India, 600036}

\author{Amareshwara Sainadh Chamarthi\footnote{Postdoctoral Researcher, California Institute of Technology, Pasadena, USA}}
\affil{California Institute of Technology, Pasadena, USA}

\begin{document}

\maketitle

\begin{abstract}


Large-eddy simulations are performed to characterize the pressure fluctuations beneath a hypersonic boundary layer approaching a compression corner. The simulations are carried out at Mach 6.04 and an inlet momentum-thickness Reynolds number of $Re_{\theta}=4340$. The compression-corner angle is varied over $10^\circ$, $20^\circ$, $30^\circ$, and $34^\circ$ spanning attached to strongly separated regimes. The peak root-mean-square wall-pressure fluctuations intensity increases sharply with separation strength, rising by $312\%$ from R20 to R30 and a further $67\%$ from R30 to R34, with the peak located downstream of reattachment. Notably, acoustic loading increases from 140 dB in the approach flow to $\approx177$ dB downstream of reattachment in the $34^\circ$ case. Further analysis reveals that intense intermittent pressure events are concentrated near the shock foot, spatially distinct from the peak acoustic-loading region, where fluctuations are relatively sustained. With increasing interaction strength, spectral energy shifts from turbulence-dominated high frequencies to broadband low-frequency motion. Band-isolated acoustic loading maps reveal high-frequency fluctuations as the dominant contributor across the interaction zone, with low-frequency fluctuations becoming locally comparable in the R34 case near the separation region. Spatio-temporal maps of bandpass-filtered pressure fluctuations reveal downstream convecting Kelvin-Helmholtz structures and upstream propagating pressure waves near the shock foot.

\end{abstract}

\section*{Nomenclature}


{\renewcommand\arraystretch{1.0}
\noindent\begin{longtable*}{@{}l @{\quad=\quad} l@{}}

$C_f$              & skin-friction coefficient \\
$f_s$              & sampling frequency \\
$H$                & shape factor \\
$L_x, L_y, L_z$    & computational domain lengths in streamwise, wall-normal, and spanwise directions \\
$M_{\infty}$       & freestream Mach number \\
$N_x, N_y, N_z$    & number of grid points in streamwise, wall-normal, and spanwise directions \\
$p_{\infty}$       & freestream pressure \\
$q_{\infty}$       & dynamic pressure \\
$Re_{\infty}$      & freestream Reynolds number \\
$Re_{\delta_2}$    & Reynolds number based on momentum thickness and wall viscosity \\
$Re_{\theta}$      & Reynolds number based on momentum thickness and freestream viscosity \\
$Re_{\tau}$        & friction Reynolds number \\
$St_{\delta_0}$    & Strouhal number based on reference boundary-layer thickness \\
$T_r$              & recovery wall temperature \\
$u, v, w$          & instantaneous streamwise, wall-normal, and spanwise velocity components \\
$u_\tau$           & friction velocity \\
$\tilde{u}, \tilde{v}, \tilde{w}$ & Favre-averaged velocity components \\
$u_{VD}^+$ & Van Driest–transformed mean velocity \\
${u^*_i}, {(uv)}^*$  & density scaled velocity and shear stress fluctuations \\
$x, y, z$          & Cartesian coordinates \\
$x_{\mathrm{ref}}$ & reference streamwise location \\
$\delta$           & boundary-layer thickness \\
$\delta_0$         & reference boundary-layer thickness \\
$\delta^*$         & displacement thickness \\
$\mu$              & dynamic viscosity \\
$\nu$              & kinematic viscosity \\
$\omega$           & angular frequency \\
$\rho$             & density \\
$\tau_w$           & wall shear stress \\
$\theta$           & momentum thickness \\

\multicolumn{2}{@{}l}{Subscripts} \\
$i$                & inlet condition \\
$\infty$           & freestream condition \\
$w$                & wall condition \\

\multicolumn{2}{@{}l}{Superscripts and operators} \\
$\overline{(\cdot)}$ & Reynolds average \\
$\widetilde{(\cdot)}$ & Favre average \\

\end{longtable*}}

\section{Introduction}
Shock wave boundary layer interactions (SWBLIs) are highly complex flow phenomena that have attracted extensive research over the past several decades \cite{dolling2001fifty, smits2006turbulent, babinsky2011shock, gaitonde2023dynamics}. These interactions occur across both internal and external flow regions in supersonic and hypersonic vehicles, especially near transonic airfoils, supersonic inlets, and engine control surfaces, often resulting in degraded aerodynamic performance. In such configurations, SWBLI imposes an adverse pressure gradient on the incoming boundary layer and, when sufficiently strong, leads to boundary-layer detachment, forming a separation bubble. \textcolor{black}{The flow field resulting from this interaction is highly complex and inherently unsteady, spanning a broad range of frequencies that are often characterised into four distinct bands. These include (a) low-frequency content involving interaction breathing and large-scale streamwise oscillations of the separation shock foot, (b) secondary bubble oscillations at lower-mid frequencies, (c) coherent shear-layer instabilities, such as Kelvin–Helmholtz rollers at mid frequencies, and (d) incoming turbulence at high frequencies \cite{adler2020dynamics, adler2022influence, gaitonde2023dynamics}}. The characteristic frequencies associated with low-frequency unsteadiness are often one or two orders of magnitude lower than the frequency of the incoming turbulent boundary layer \cite{clemens2014low}. Such unsteady behaviour is of particular concern because it can coincide with the resonance frequencies of flight structures, potentially leading to structural failure \cite{mcnamara2011aeroelastic}. 

Early computational studies of high-speed flows have established the behaviour of wall pressure fluctuations in the turbulent boundary layer \cite{duan2016pressure, ritos2017implicit, ritos2019acoustic, zhao2021wall, drikakis2021flow}. In particular, Ritos et al. \cite{ritos2019acoustic} studied the acoustic loading beneath the hypersonic transitional and turbulent boundary layers across a wide range of Mach numbers. The authors concluded that the transition region experiences the most severe acoustic load, with large-scale vortical structures causing pressure fluctuations that are nearly 38\% ($\approx3$ dB) higher than in the fully turbulent region. In addition, an increase in the intensity of the free-stream turbulence is observed to amplify the pressure fluctuations in the fully turbulent region. Similar experimental and computational studies by Zhao and Zhao \cite{zhao2021wall} and Drikakis et al. \cite{drikakis2021flow} confirmed that the boundary-layer transition significantly increases acoustic loading levels in hypersonic flows. Drikakis et al. \cite{drikakis2021flow} further reported differences of up to 5 dB between the implicit large eddy simulation (iLES) and the significantly more refined direct numerical simulation ($\times50$) in the transition region, reinforcing the need for high-fidelity simulations for reliable acoustic-loading predictions.

On the other hand, SWBLI interactions substantially modify the wall-pressure field relative to undisturbed boundary layers. For instance, Bernardini et al. \cite{bernardini2011wall} examined the signature of wall pressure fluctuations in a $M_{\infty}= 1.3$ transonic SWBLI. The results showed that the interaction region strongly amplifies pressure fluctuations, \textcolor{black}{with root-mean-square wall pressure fluctuations exceeding 162 dB, an increase of 7 dB relative to the fluctuations in the incoming turbulent boundary layer.} Whalen et al. \cite{whalen2019unsteady} reported a similar amplification trend in the ramp-induced transitional and turbulent SWBLI at Mach 6, where the reattachment region showed the strongest root-mean-square (RMS) pressure. Furthermore, Ritos et al. \cite{ritos2020computational} quantified the aeroacoustic loading in Mach 7 transitional and shock-induced turbulent boundary-layer flows. The authors reported that $p_{rms}$ levels increased from approximately 140 dB in the upstream TBL to more than 170 dB downstream of reattachment, with extremely high values coinciding with regions of maximum pressure gradient. Further insight into SWBLI loading characteristics is provided by compression–decompression \cite{duan2021direct} and conical interaction studies \cite{zuo2021reynolds, zuo2023wall}, which show that the most intense load is concentrated in the reattached boundary layer and the mean shock region. 


The aforementioned studies indicate that the effects of hypersonic SWBLI on acoustic loading remain insufficiently explored. To facilitate the development of a hypersonic aircraft design, characterising wall-pressure fluctuations beneath a hypersonic boundary layer is crucial. This characterisation enables an accurate prediction of aeroacoustic loading and assessment of structural fatigue. Conventional eddy-viscosity-based RANS models are time-averaged and cannot capture these unsteady loads, and instead rely on a semi-empirical approach to estimate acoustic loads. \textcolor{black}{Notably, Ritos et al. \cite{ritos2019wall} addressed the limitations of existing wall-pressure spectra models for supersonic and hypersonic flows by introducing compressibility corrections. Following this, Kaluva et al. \cite {kaluva2023framework} later developed a RANS-based predictive framework to estimate acoustic loads in both attached turbulent boundary layers and the SWBLI. Despite the recent progress, acoustic prediction models are not generic enough and remain under development.} In contrast, eddy-resolving approaches such as DNS and LES can accurately quantify unsteady loads, provide deeper physical insight into SWBLI, and support the development of low-order acoustic prediction models. Motivated by these observations, we aim to investigate a hypersonic boundary layer approaching a compression corner at different ramp angles using high-fidelity LES. \textcolor{black}{Although such configurations have been experimentally investigated by Whalen et al. \cite{whalen2019unsteady}, the focus has primarily been on the flow structure and shock dynamics. Consequently, the acoustic loading characteristics across hypersonic compression-corner interactions remains limited. The present study addresses the following questions: (a) How does the acoustic loading environment evolves as the interaction transitions from attached to strongly separated regimes with increasing ramp angle? (b) Does the resulting acoustic loading exhibit relatively uniform or intermittent characteristics? (c) How do distinct frequency bands contribute to the total acoustic loading, and does their spatial dominance across the interaction region change systematically with interaction strength?}

The paper is organised as follows: Section \ref{numerics} discusses the numerical framework and the case setup. The results are then presented in Section \ref{results}, starting with the numerical sensitivity to grid resolution and validation of the incoming turbulent boundary layer. This is followed by a discussion on SWBLI flow organisation and mean flow characteristics, together with the characterisation of wall-pressure fluctuations, including their spectral and band-resolved features. Section \ref{conclusion} summarizes the main findings.

\section{Numerical Methodology} \label{numerics}

\subsection{Numerical Framework}

A high-order finite-difference in-house solver, COMPSQUARE \cite{vadlamani2018distributed}, is employed to carry out all the simulations presented in this study. The unsteady three-dimensional compressible Navier-Stokes equations are solved in a non-dimensional, conservative form using a generalised curvilinear coordinate framework. The solver is parallelised using OpenACC and OpenMPI directives, enabling GPU acceleration and scalable multi-block computations. The in-house solver has been thoroughly validated across a variety of benchmark test cases, including supersonic turbulent boundary layers over a flat plate and shock-wave/turbulent boundary-layer interactions in both supersonic and hypersonic regimes \cite{kumar2022scalar, kumar2025assessment}. Strong flow discontinuities in high-speed turbulent regimes are captured using a centralised gradient-based reconstruction (C-GBR) scheme \cite{hoffmann2024centralized}. In the current formulation, conservative variables are stored at node locations, and numerical convective fluxes are evaluated at cell interfaces using finite-difference approximations. Spatial derivatives of the numerical fluxes are subsequently reconstructed at the nodes using the fluxes value at the cell interfaces, yielding a conservative finite-difference formulation. A sixth-order monotonicity-preserving explicit scheme (MEG6) is used for spatial discretisation, while the solution is marched in time using a third-order total-variation-diminishing Runge–Kutta scheme (RK-3). The convective numerical fluxes are estimated using the HLLC (Harten-Lax-van Leer-Contact) approximate Riemann solver. The numerical viscous fluxes and their spatial derivatives are evaluated at the node locations. 

\textcolor{black}{The free-stream temperature in the present simulations is $T_{\infty} = 63.6$ K. At such a low ambient temperature, Priebe and Martin \cite{priebe2021turbulence} recommended using Keyes’ law over Sutherland's law for dynamic viscosity calculation. Accordingly, the dynamic viscosity is computed using Keyes’ law as  $\mu = 1.488 \times 10^{-6}\sqrt{T}/{(1 + \left(122.1/T\right) \times 10^{-5/T})}$, where $\mu$ is in Pa s and $T$ is the temperature in Kelvin.} A digital-filter-based turbulence generator \cite{veloudis2007novel, touber2009large} is employed at the inflow to prescribe physically realistic turbulent fluctuations. To account for wall-normal variations in turbulence scales, the inflow plane is divided into two regions. Each region is prescribed distinct integral length scales for the velocity components ($u$, $v$, $w$) based on the prior studies reported in the literature \cite{kumar2025assessment}. The thermodynamic fluctuations are introduced at the inflow plane using the Strong Reynolds Analogy (SRA) \cite{touber2009large}.  Velocities are normalised using $u_\infty$, while density, pressure, and temperature are normalised by $\rho_\infty$, $\rho_\infty u_\infty^2$, and $T_{\infty}$, respectively. \textcolor{black}{All simulations are performed using implicit large-eddy simulation (iLES) without an explicit subgrid-scale (SGS) model. Instead, the effects of unresolved scales are represented implicitly through the inherent numerical dissipation of the MEG6 spatial discretisation scheme. \cite{chandravamsi2023application}.}

\begin{table}
\caption{Freestream and wall boundary conditions for the present simulations.}
\label{tab:flow_properties}
\centering
\begin{tabular}{cccccccc}
\hline
\hline
$M_{\infty}$ &
$\mathrm{Re}_{\infty}/\mathrm{m}$ &
$p_{\infty}$ (kPa) &
$T_{\infty}$ (K) &
$\rho_{\infty}$ (kg\,m$^{-3}$) &
$u_{\infty}$ (m\,s$^{-1}$) &
$T_{w}$ (K) &
$T_{w}/T_{r}$ \\
\hline
6.04 &
$23.6\times10^{6}$ &
2.04 &
63.6 &
0.113 &
962 &
300 &
0.625 \\
\hline
\hline
\end{tabular}
\end{table}

\begin{figure}
    \centering
    \includegraphics[scale=0.95]{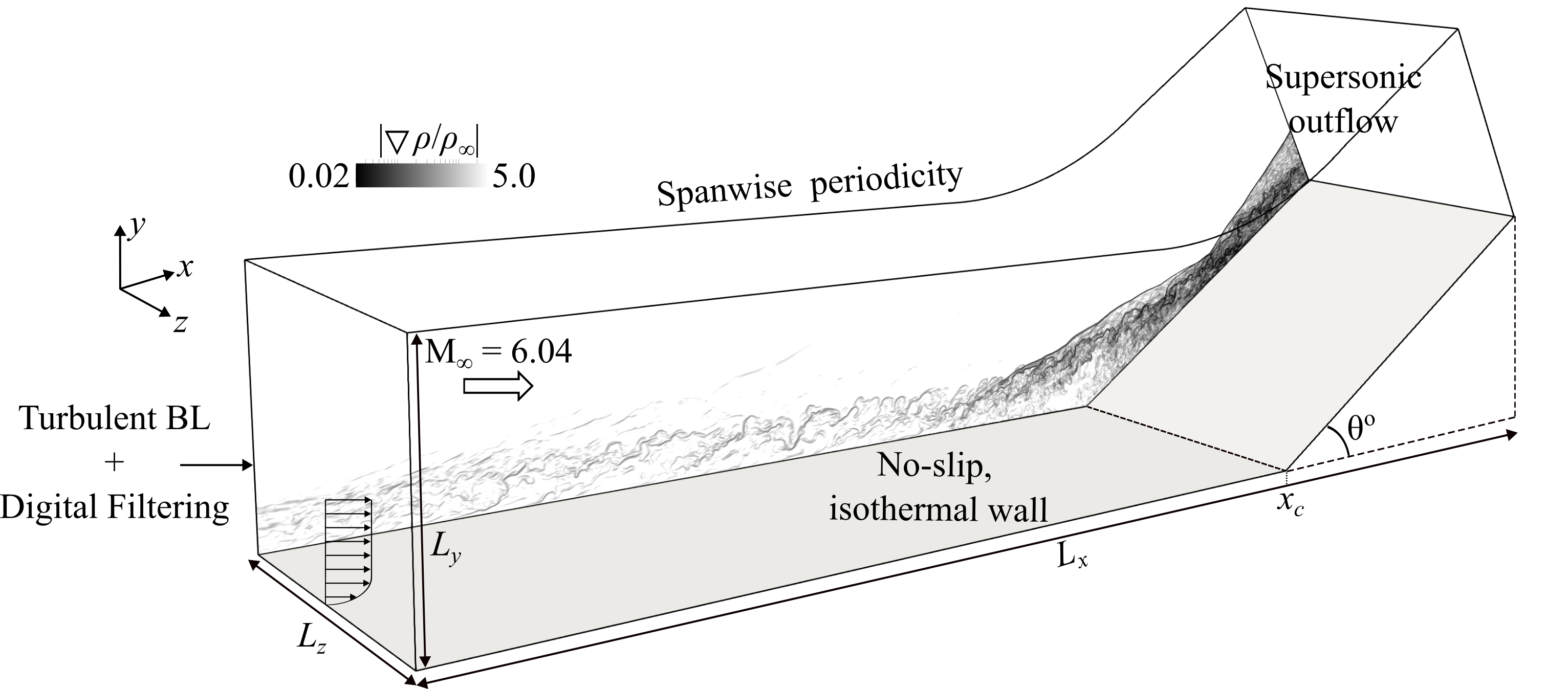}
    \caption{Schematic illustrating the computational domain along with the boundary conditions. The contours of density gradient magnitude are shown on the $xy$ plane.}
    \label{fig:schematic}
\end{figure}

\subsection{Flow configuration and computational setup} \label{setup}

We investigate the effect of varying compression ramp corners on a hypersonic turbulent boundary layer at a free-stream Mach number of $M_{\infty} = 6.04$. Four different compression corner angles are simulated, namely $10^{\circ}, 20^{\circ}, 30^{\circ}$ and $34^{\circ}$. The flow conditions are in line with the experimental campaign carried out by Whalen et al. \cite{whalen2019unsteady} and are listed in Table \ref{tab:flow_properties}. The Reynolds number based on the inlet momentum thickness ($\theta_{\text{in}}$) and the free-stream viscosity is $Re_{{\theta}_{\text{in}}}$ = 4340.  Figure \ref{fig:schematic} shows the schematic of the computational domain along with the boundary conditions used for the present study. \textcolor{black}{The LES domain is adapted from the RANS study of Kaluva et al. \cite{kaluva2023framework} and is truncated to retain only the region of interest for the high-fidelity simulations.
In addition, the computational domain is configured to reproduce a boundary-layer thickness of $\delta_0=5.25$ mm at $\approx$ 18 mm upstream of the compression corner, in accordance with the experimentally measured value of $\delta_0 = 5.7 \pm 0.45$ mm reported by Whalen et al. \cite{whalen2019unsteady, whalen2020hypersonic}}. The computational domain extends $L_x/\theta_{\text{in}}\approx750.5-789.2$ in streamwise direction, $L_y/\theta_{\text{in}}\approx143$ in wall-normal direction, and  $L_z/\theta_{\text{in}}=216$ in spanwise 
direction, where the inlet momentum thickness $\theta_{\text{in}}=0.184 ~\text{mm}$. The streamwise distance from the inlet to the compression corner and the ramp length are kept constant for all cases, resulting in different streamwise extents for different compression corner angles. The turbulent fluctuations generated using the digital filtering method, along with the mean turbulent boundary layer (TBL) profiles, are imposed at the inflow plane. \textcolor{black}{Table \ref{tab:df_scales} summarises the multi-zone integral length scales employed for the present study \cite{kumar2025assessment}.} \textcolor{black}{The integral length scales in the streamwise, wall-normal and spanwise directions are denoted as $I_x$, $I_y$, and $I_z$ 
respectively. The displacement thickness at the inflow boundary, $\delta_1^{\text{vd}}$, is computed from the van Driest–transformed velocity profile using the incompressible-flow definition. It is worth noting that while introducing} thermodynamic fluctuations at the inlet using SRA, significant density and temperature fluctuations (both positive and negative) are observed at the inlet, leading to numerical instability. Consequently, the thermodynamic fluctuations are introduced using the SRA approach at a lower Mach number ($M \approx 4$), allowing physically realistic fluctuations to develop naturally within the computational domain. The flow variables are extrapolated at the top and the outlet plane, while periodicity is enforced in the spanwise direction. \textcolor{black}{A no-slip, isothermal wall boundary condition is specified at the bottom wall with $T_w/T_r = 0.625$, where the recovery temperature is computed as $T_r=1+0.5r(\gamma-1)M_{\infty}^2$ and $r=0.89$.}

\begin{table}[!t]
\centering
\caption{Integral length scales used for digital filtering.}
\label{tab:df_scales}
\begin{tabular}{lccc}
\hline
\hline
 & $u$ & $v$ & $w$ \\
\hline
$I_x/\delta_0$ & 0.7 & 0.28 & 0.28 \\

$I_y/\delta_0$ 
& $0.13^{\mathrm{inn}} - 0.23^{\mathrm{out}}$
& $0.17^{\mathrm{inn}} - 0.30^{\mathrm{out}}$
& $0.10^{\mathrm{inn}} - 0.13^{\mathrm{out}}$ \\

$I_z/\delta_0$ & 0.15 & 0.15 & 0.30 \\
\hline
\hline
\end{tabular}

\vspace{1mm}

\footnotesize
\textit{Note:} $\mathrm{inn}$ corresponds to $y \le y_{\mathrm{lim}}$ and 
$\mathrm{out}$ corresponds to $y > y_{\mathrm{lim}}$, where 
$y_{\mathrm{lim}} = \delta_1^{\mathrm{vd}}$.
\end{table}

Table \ref{tab:grid_sensitivity} summarises the grid sensitivity parameters for the $34^{\circ}$ (R34) case, chosen as the reference case owing to its stronger shock interaction and larger separation bubble. The inner-scale grid resolution is computed as $\Delta x^+ = u_{\tau} \Delta x / \nu_{w}$, where $u_{\tau}$ is the friction velocity and $\nu_{w}$ is the wall kinematic viscosity. The minimum and the maximum values of streamwise grid resolution ($\Delta x^+$) correspond to the compression corner and the reference upstream location $x_{\text{ref}}=-3.5\delta_0$, respectively. The coarser mesh (G1) comprises 270.4 million grid points, whereas the finer mesh (G2) comprises 1.03 billion grid points. Additionally, the statistical analysis employs two types of averaging. The Reynolds average of a quantity $\varphi$ is denoted as $\overline{\varphi}$, and the corresponding fluctuation is defined as $\varphi' = \varphi - \overline{\varphi}$. On the other hand, Favre-averaged of a quantity $\varphi$ is denoted as $\tilde{\varphi}$, and the corresponding Favre fluctuation is defined as $\varphi'' = \varphi - \tilde{\varphi}$. For each case, statistical averages are calculated over 10 flow-through times (FTT), where each FTT represents the time required for a fluid parcel to travel from the inflow plane to the domain outlet. The resulting mean boundary layer profiles and Reynolds stress components are confirmed to be statistically converged. 

\begin{table*}[htpb]
\caption{Grid sensitivity study for the R34 ramp case}
\label{tab:grid_sensitivity}
\centering
\setlength{\tabcolsep}{9pt}
\renewcommand{\arraystretch}{1.2}
\begin{tabular}{lcccc}
\hline
\hline
Case &
\begin{tabular}[c]{@{}c@{}}No. of points \\ ($N_x \times N_y \times N_z$)\end{tabular} &
$\Delta x^{+}_{\min,\max}$ &
$\Delta y^{+}_{\min}$ &
$\Delta z^{+}$ \\
\hline
R34-G1  & $1561\times330\times525$ & 4.8 -- 7.8 & 0.94 & 4.9 \\
R34-G2  & $3218\times400\times801$ & 1.9 -- 4.0 & 0.94 & 3.3 \\
\hline
\hline
\end{tabular}
\end{table*}

\section{Results} \label{results}

\subsection{Sensitivity to Grid Resolution and Adequacy of Spanwise extent} \label{subs:sensitivity_studies}

\begin{figure}
    \centering
    \includegraphics{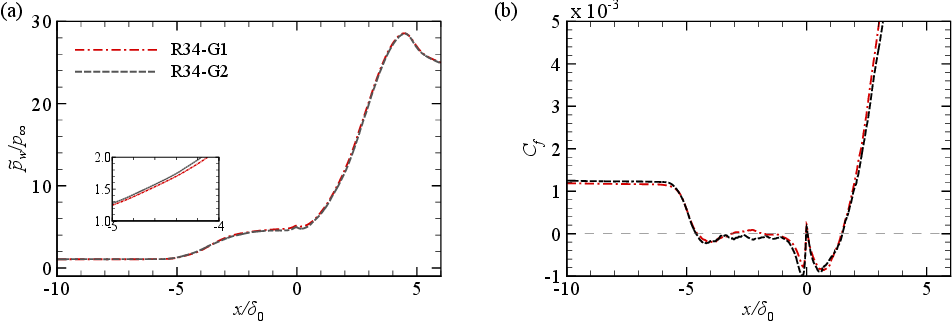}
    \caption{Streamwise variation of mean (a) wall pressure, and (b) skin friction coefficient for R34 case across different mesh resolutions.}
    \label{fig:grid_sensi}
\end{figure}

The influence of the grid resolution on the compression-corner interaction is examined. Figures \ref{fig:grid_sensi} (a) and (b) compare the streamwise evolution of mean wall pressure and skin friction coefficient for G1 and G2 mesh resolutions. The streamwise coordinate is normalised using $\delta_0$, with the compression corner as the origin. As observed from the $C_f$ distribution, the G2 mesh over-predicts the extent of the separation bubble by only $0.5\%$ compared to the G1 mesh. Moreover, the results show that the predictions for $\tilde{p}_w$ and $C_f$ are almost identical across both mesh resolutions, indicating grid convergence at this resolution. Henceforth, all simulations and corresponding discussions are based on the G1 grid resolution. 

\begin{figure}
    \centering
    \includegraphics[scale=1.0]{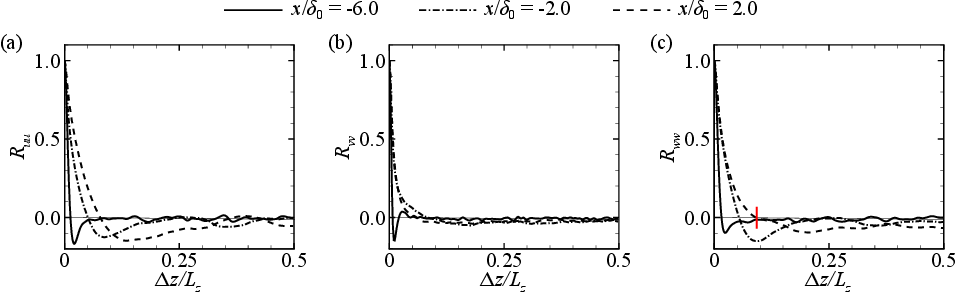}
    \caption{Two-point spanwise autocorrelation velocity functions computed at fixed $y/\delta_0=0.1$ with (a) $x/\delta_0$ = -6.0, (b) $x/\delta_0$ = -2.0, and (c) $x/\delta_0$ = 2.  }
    \label{fig:span_autocorr}
\end{figure}

\textcolor{black}{To assess the adequacy of the considered spanwise extent of the domain, a spanwise autocorrelation analysis is performed.} Figure \ref{fig:span_autocorr} shows the autocorrelation of velocity fluctuations computed along the homogenous spanwise direction at three different streamwise locations at a fixed $y/\delta_0 = 0.1$ for the R34 case. \textcolor{black}{This test case is selected because it exhibits the largest unsteadiness due to its stronger interaction strength and correspondingly larger separation bubble. Demonstrating spanwise adequacy for this case ensures that the domain extent is also sufficient for the lower interaction-strength cases.} These  include the undisturbed TBL ($x/\delta_0=-6$), the midpoint of the mean separation bubble ($x/\delta_0=-2$), and the region downstream of the reattachment ($x/\delta_0=2$).  In the upstream boundary layer, the streamwise correlation function ($R_{uu}$) exhibits the characteristic turbulent boundary layer signature, with negative values at larger separations, which are caused by alternating low- and high-speed streaks \cite{laguarda2024reynolds}. The wall-normal velocity correlation function ($R_{vv}$) decays to zero much more rapidly than the streamwise and spanwise components. Among all the locations considered, the upstream TBL exhibits the shortest spanwise coherence length, which is consistent across all the velocity-fluctuation components. An increase in the levels of coherence, as indicated by the first zero crossing, is observed at the centre of the mean separation bubble (Fig. \ref{fig:span_autocorr} (b)). At this location, while streamwise and spanwise velocity fluctuations approach zero correlation by $0.25L_z$, wall-normal fluctuations decorrelate over a much shorter spanwise distance. Downstream of the reattachment, a further increase in coherence length is observed in comparison to the upstream TBL and the central separation region \textcolor{black}{(indicated with red solid line)}. Moreover, the streamwise and the spanwise velocity fluctuations remain weakly correlated, with the correlation values of $\approx0.05$ at $0.5L_z$. These observations indicate that the selected spanwise extent is adequate to resolve the dominant spanwise coherence scales for the present simulations.

\subsection{Validation of Incoming TBL}

\begin{figure}[t]
    \centering
    \includegraphics{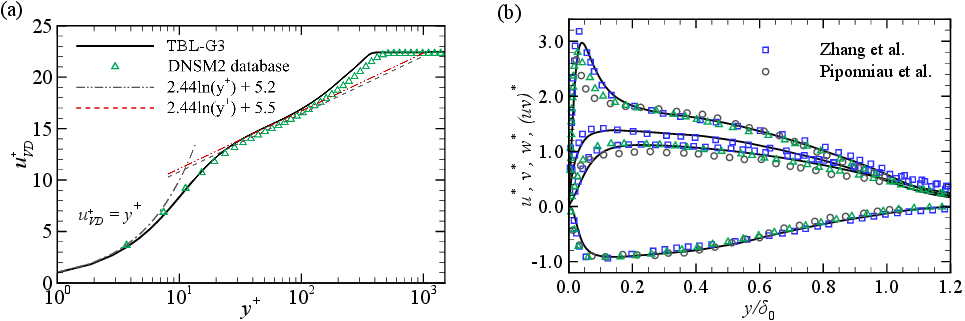}
    \caption{Plots of (a) mean Van Driest transformed mean velocity and (b) density scaled velocity and shear stress fluctuations at $x = -3.5\delta_0$. Symbols denote data from DNSM2 databse \cite{DNSM2} (triangles), Piponniau et al. \cite{piponniau2009simple} (squares), and Zhang et al. \cite{Zhang2024EffectsWallTemperature} (circles).}
    \label{fig:stats1}
\end{figure}

The state of the upstream turbulent boundary layer (TBL) is examined and validated against reference numerical simulations and established correlations reported in the literature. To this end, additional simulations are performed in the absence of the compression-ramp corner using the same flow configuration and computational setup described in section \ref{setup}. The boundary layer statistics are obtained at a reference location of $x_{\mathrm{ref}}=-3.5\delta_0$. The reference location corresponds to the experimental measurement position ($\approx$ 18 mm upstream of the compression corner) , where the boundary-layer thickness is measured \cite{whalen2019unsteady}. Table \ref{tab:tbl_parameters} lists the characteristics of the turbulent boundary layer at $x_{\mathrm{ref}}$, including the ratio of reference to the inlet boundary layer thickness  ($\delta_0/\delta_{i}$) and \textcolor{black}{Reynolds number defined based on different characteristic length scales and viscosity definitions, namely $Re_\theta$, $Re_{\delta_2}$, and $Re_\tau$.} Figures \ref{fig:stats1} (a) plots the mean Van Driest transformed velocity (${u}_\mathrm{VD}^+$), which accounts for compressibility effects through scaling of the mean velocity gradient with $\sqrt{\bar{\rho}/\bar{\rho}_w}$, where $\overline{\rho}_w$ is the mean wall density \cite{van1951turbulent}. The transformation is computed as

\begin{equation}
{u}^{+}_{VD} = \int_{0}^{\bar{u}^{+}} \sqrt{\bar{\rho}/\bar{\rho}_w} \, d\bar{u}^{+}
\end{equation}

\textcolor{black}{where, $\overline{\rho}$ is the Reynolds averaged density.} The LES data is also compared with the DNS profiles ($Re_{\delta_2}=1384$ and $Re_{\tau}=402$) from the DNSM2 database \cite{DNSM2}. The predictions of ${u}_{\mathrm{VD}}^{+}$ show good agreement with the reference data, although marginal deviations are observed in the logarithmic region. These deviations arise from differences in wall thermal conditions, as the current simulations utilise a cooled wall ($T_w/T_r \approx 0.63$), whereas the reference simulations are carried out with an adiabatic wall. Moreover, the van Driest transformation is known to be most effective for adiabatic or quasi-adiabatic walls. The viscous sublayer exhibits linear collapse up to $(y^{+}\approx 3 $), after which deviations from incompressible behaviour are observed. The logarithmic region in incompressible TBL is characterised with a slope, $1/\kappa \approx 2.44$ and an intercept of $C \approx 5.2$. In the present simulation, the intercept is slightly higher, with $C \approx 5.5$. Elevated intercept values have also been reported in previous studies of hypersonic TBL, with $C=5.9$ at Mach 7.2 \cite{priebe2021turbulence}, $C=5.2-5.6$ across a wide range of Mach numbers $(M_{\infty} = 2,5, \ \text{and} \ 11)$ \cite{huang2022direct}. 

To account for the compressibility effects, the density-scaled velocity, ${u}_i^{*}$ and shear stress fluctuations, $({uv})^{*}$ are computed as
\begin{equation}
u_i^* = \frac{\sqrt{\bar{\rho}/\bar{\rho}_w}}{u_\tau} \sqrt{\widetilde{u_i''^2}}, 
\qquad
(u_i u_j)^* = \frac{\bar{\rho}/\bar{\rho}_w}{u_\tau^2} \, \widetilde{u_i'' u_j''}.
\end{equation}

\begin{figure}[!t]
    \centering
    \includegraphics{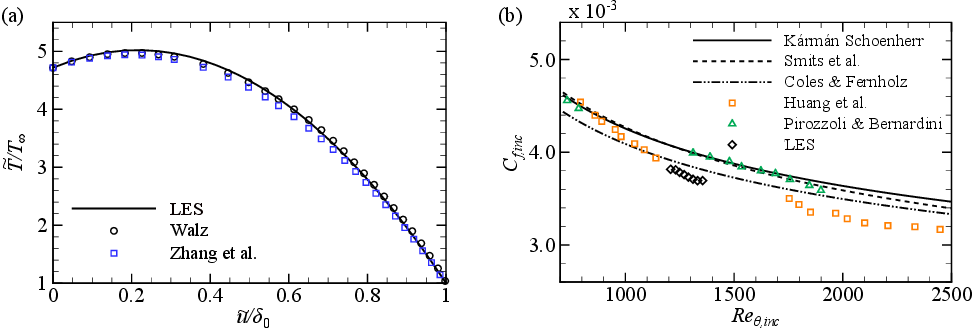}
    \caption{Plots of (a) temperature-velocity relations at $x = -3.5\delta_0$ and (b) Van Driest II transformed skin friction coefficient compared against the correlations of Kármán–Schoenherr \cite{roy2006review}, Smits et al. \cite{smits1983low}, and Coles \& Fernholz \cite{nagib2007approach}, and (b)  Symbols denote data by Pirozzoli and Bernardini \cite{pirozzoli2011turbulence} (triangles), Huang et al. \cite{huang2022direct} (orange squares), Walz equation \cite{walz1969boundary} (circles), and generalised relation of Zhang et al. \cite{zhang2014generalized} (blue squares).}
    \label{fig:stats2}
\end{figure}

Figure \ref{fig:stats1} (b) compares the density-scaled velocity and shear stress fluctuations against the reference studies reported in the literature \cite{piponniau2009simple, DNSM2, Zhang2024EffectsWallTemperature}. The predicted fluctuation trend is in favourable agreement with the reference data. The marginal differences in the amplitude and location of the peak streamwise velocity fluctuations (${u}^{*}$) are in agreement with the Reynolds-number dependence reported by Pirozzoli and Bernardini \cite{pirozzoli2011turbulence}. Overall, this agreement confirms that the upstream TBL is well captured. 
Figure \ref{fig:stats2} (a) compares the mean temperature and streamwise velocity relation at $x=-3.5\delta_0$ with the empirical correlations proposed by Walz \cite{walz1969boundary}, and Zhang et al. \cite{zhang2014generalized}. \textcolor{black}{The predictions of LES compare favourably with the empirical correlations. The increase in the near-wall temperature-velocity relation is characteristic of strong wall-cooling effects.} Figure \ref{fig:stats2} (b) compares the mean Van Driest II transformed skin-friction coefficient with the correlations of Kármán–Schoenherr \cite{roy2006review}, Smits et al. \cite{smits1983low}, and Coles \& Fernholz \cite{nagib2007approach} alongside the DNS data of Huang et al. \cite{huang2022direct}. A Van Driest II \cite{van1956turbulent} transformation is used to relate compressible boundary-layer skin friction to an equivalent incompressible value. As shown in the figure, $C_{f,inc}$ and $Re_{\theta,inc}$ represent the equivalent incompressible skin-friction coefficient and the incompressible Reynolds number based on momentum thickness, respectively. The LES data of $C_{f,inc}$ spans over the streamwise extent $x=-12\delta_0$ to $-3.5\delta_0$. \textcolor{black}{The prediction shows a deviation of $-7.2\%$ to $-6.2\%$, in line with the findings of Huang et al. \cite{huang2022direct}, where deviations between $-15\%$ and $-6\%$ for hypersonic cold-wall cases were reported.} The present LES shows good agreement with both the empirical models, thereby validating the current framework for the upstream turbulent boundary layer.

\begin{table}[t]
\caption{Turbulent boundary-layer parameters at $x=-3.5\delta_0$, where the Reynolds numbers are defined as ${Re_{\theta}} = \rho_\infty u_\infty \theta / \mu_{\infty}$
${Re}_{\delta_2} = \rho_\infty u_\infty \theta / \mu_w$ and 
${Re}_\tau = \rho_w u_\tau \delta / \mu_w$ with $u_\tau$ as the friction velocity.}
\label{tab:tbl_parameters}
\centering
\begin{tabular}{ccccccc}
\hline\hline
Case &
$\delta_0/\delta_{i}$ &
$\delta_0/\theta$ &
$H (\delta^*/\theta)$ &
$Re_\theta$ &
$Re_{\delta_2}$ &
$Re_{\tau}$ \\
\hline
TBL & 1.23 & 23.53 & 12.48  & 5530 & 1362 & 341 \\
\hline\hline
\end{tabular}
\end{table}

\subsection{Flow organisation and mean flow characteristics}

\begin{figure}[!t]
    \centering
    \includegraphics{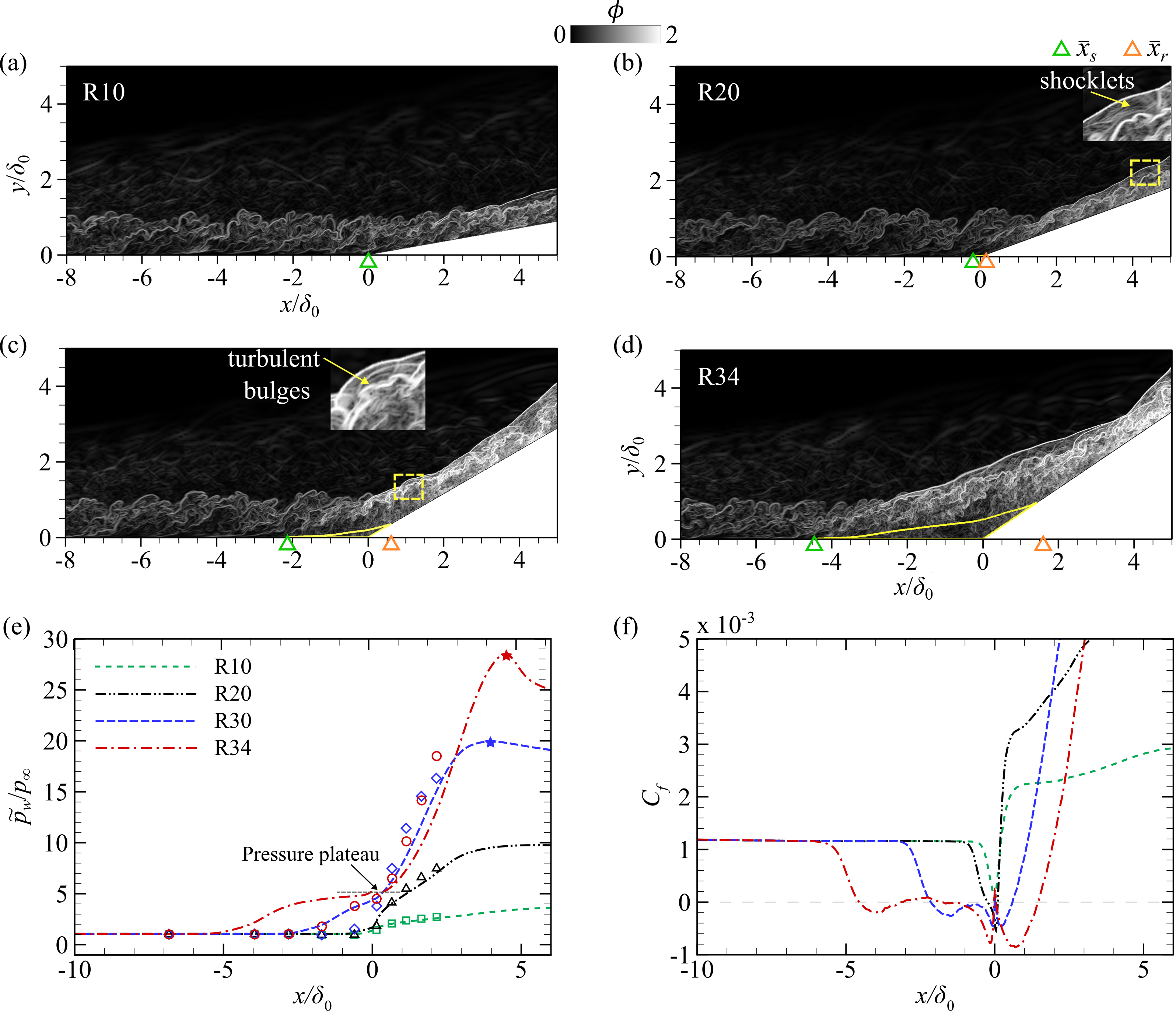}
    \caption{\textcolor{black}{Instantaneous contours of the logarithmically scaled density gradient ($\phi=\text{log}(|\nabla \rho| + 1)$) at the midspan location for cases of (a)  R10, (b) R20, (c) R30, and (d) R34. The triangle symbols indicate mean separation (green) and reattachment (orange) locations. Plots of mean (e) wall pressure and (f) skin friction coefficient along the streamwise direction for all the ramp angles. Symbols indicate the experimental data of Whalen et al. \cite{whalen2019unsteady}.}}
    \label{fig:grad_rho}
\end{figure}

Following validation of the upstream turbulent boundary layer, the present section explores the effect of compression corner angles on the resulting SWBLI characteristics. \textcolor{black}{Figures \ref{fig:grad_rho} (a)--(d) show the instantaneous contours of logarithmically scaled density gradient ($\phi=\text{log}(|\nabla \rho| + 1)$) at the midspan location, while the streamwise evolution of the mean wall pressure and skin friction coefficient ($C_f=2\tau_w/(\rho_{\infty} u_{\infty}^2)$) are shown in Fig. \ref{fig:grad_rho} (e) and (f) respectively, for all the ramp angles. The triangle symbol denotes the mean separation and the reattachment locations, while the solid lines (yellow) indicate the mean separation region. A magnified inset of the highlighted region is also shown.} In all ramp cases, the turbulent structures downstream of the interaction become more pronounced and brighter, indicating enhanced fluctuations following the shock interaction and reattachment. As observed with the small ramp angle (R10), the associated pressure rise is often insufficient to induce upstream boundary-layer separation, and the flow remains attached. With an increase in the ramp angle (R20), the stronger adverse pressure gradient promotes the formation of a separation bubble, accompanied by an oblique shock that shifts upstream of the ramp corner, marking the onset of incipient separation \cite{settles1979detailed}. This behaviour is consistent with the $C_f$ distribution (Fig. \ref{fig:grad_rho} f), which shows a progressive increase in separation with increasing ramp angle. For the attached and incipient cases, $C_f$ remains positive upstream of the compression corner and exhibits only a localised dip near the compression corner, followed by a rapid downstream recovery. With a further increase in ramp angle (R30 and R34), the separation bubble enlarges, and the shock system moves progressively farther upstream. These observations are further supported by the $C_f$ distribution, in which the onset of $C_f$ shifts further upstream, while its recovery to positive values occurs downstream, indicating a larger recirculation bubble.  The global minimum $C_f$ also becomes increasingly negative, aligned with stronger reverse flow in the separated region. Moreover, the contours indicate a slight distortion of the shock system induced by turbulent bulges in the upstream boundary layer. Localised shocklets form over these bulges, which become more prominent in the separated cases. In addition, weak oblique shocks emanating from the inlet are visible as a consequence of the finite incoming boundary-layer thickness, accompanied by weak compression waves that gradually decay in the streamwise direction. Table \ref{tab:Separation_param} summarizes the mean separation and reattachment locations together with the corresponding separation and interaction lengths. Following Souverein et al. \cite{souverein2013scaling}, $L_{\text{int}}$ is defined as the distance from the compression corner to the upstream location where $p_w(x) \ge 1.02\,p_{w,\mathrm{u}}$, with $p_{w,\mathrm{u}}$ denoting the wall pressure in the undisturbed turbulent boundary layer. The tabulated values confirm the trends observed in the skin-friction distributions.

Figure \ref{fig:grad_rho} (e) compares the streamwise evolution of the mean wall pressure for all the ramp angles. The experimental results of Whalen et al. \cite{whalen2019unsteady} are also overlaid. With increasing ramp angle, the onset of wall-pressure rise shifts progressively upstream. In the attached (R10) and incipient-separation cases (R20), the wall pressure increases monotonically through the interaction before gradually approaching a downstream plateau, indicating recovery of the pressure jump. In contrast, the largely separated cases (R30 and R34) exhibit an earlier upstream pressure rise and a pronounced pressure plateau at the compression ramp, characteristic of the separated region. Further downstream of the compression corner, the pressure continues to rise, peaks (indicated by an asterisk), and then decreases, indicating a local transition from an adverse to a favourable pressure gradient. \textcolor{black}{The pressure peak is attributed to strong compression during reattachment in the separated hypersonic interaction, resulting in wall pressure exceeding the inviscid oblique-shock level. Similar wall-pressure behaviour has been reported by Zhang et al. \cite{Zhang2024EffectsWallTemperature}.} The experimental measurements of Whalen et al. \cite{whalen2019unsteady} indicate the onset of incipient separation at \(30^\circ\), followed by fully developed separation at \(34^\circ\). In contrast, the LES predicts attached flow at \(10^\circ\) and the onset of incipient separation starting from \(20^\circ\). While the wall-pressure distributions for the attached (R10) and the incipient (R20) cases remain in favourable agreement with the reference data, the largely separated cases (R30 and R34) show an upstream pressure rise together with a slight under-prediction of the pressure downstream of the compression corner, indicating a modest overestimation of the separation bubble size. Since the LES setup is based on the experimentally prescribed boundary-layer thickness at $x_{\text{ref}}$, the upstream conditions are consistent with those of the reference case. The good agreement between R10 and R20 suggests that the discrepancy becomes significant only for cases that are largely separated. Furthermore, the grid-sensitivity study for the R34 case confirms mesh convergence, making insufficient spatial resolution an unlikely cause. \textcolor{black}{Although the origin of the discrepancy between the simulation and the experiment is unclear, a more plausible explanation is the influence of three-dimensional effects arising from the finite span of the experimental ramp plate, as the plate does not extend across the full test-section width. Such finite-span effects are known to alter both the location and extent of the separation bubble as reported by Bruce et al. \cite{bruce2011corner} and Volpiani et al. \cite{volpiani2020effects} and therefore, may explain a relatively larger separation length predicted in the present simulations.} 


\begin{table}[t]
\caption{Streamwise positions of interest with separation length.}
\label{tab:Separation_param}
\centering
\begin{tabular}{ccccc}
\hline\hline
Case &
$\overline{x}_{s}/\delta_{0}$ &
$\overline{x}_{r}/\delta_{0}$ &
$\overline{L}_{sep}/\delta_{0}$ &
$\overline{L}_{int}/\delta_{0}$ \\
\hline
R10 & 0.0 & 0.01 & 0.01  & 0.41  \\
R20 & -0.23 & 0.10 & 0.33  & 0.81  \\
R30 & -2.14 & 0.60 & 2.74  & 2.92  \\
R34 & -4.54 & 1.45 &  6.0 & 5.54  \\
\hline\hline
\end{tabular}
\end{table}
\subsection{Characterisation of Wall-Pressure Fluctuations}

\begin{figure}
    \centering
    \includegraphics{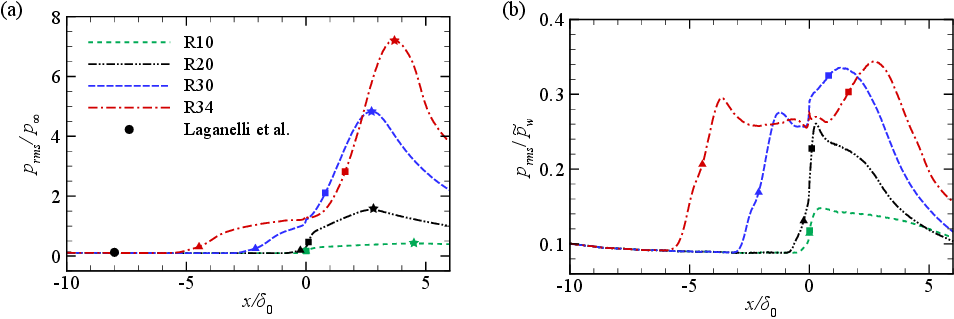}
    \caption{Streamwise distribution of root mean square wall pressure fluctuations ($p_{rms}$) normalized with (a) freestream pressure, and (b) local mean wall pressure. The filled symbols indicate mean separation (triangle) and reattachment (square) positions. The theoretical prediction from the Laganelli model \cite{laganelli1983wall} is indicated by a black-filled circle.}
    \label{fig:prms_plots}
\end{figure}


Figures \ref{fig:prms_plots} (a) and (b) show the streamwise evolution of  $p_{\mathrm{rms}}$ normalized by $p_\infty$ and $\tilde{p}_w$, respectively. The mean separation and reattachment locations are also overlaid. Unlike the mean wall pressure, where a pronounced peak is observed only for R30 and R34 (Fig. \ref{fig:grad_rho} e), all four cases exhibit a distinct $p_{\mathrm{rms}}$ peak (indicated with an asterisk symbol), located downstream of reattachment for the separated cases and downstream of the corner for the attached case. The post-reattachment $p_{rms}$ peak is attributed to the impingement of energetic unsteady vortical structures within the reattaching shear layer. For the fully separated cases, this peak is shifted slightly upstream relative to the mean-pressure peak by $\approx1.1\delta_0$ for R30 and $\approx0.8\delta_0$ for R34. The fluctuation levels increase substantially with ramp angle, with the peak value rising by approximately $312\%$ from R20 to R30 and by a further $67\%$ from R30 to R34. Unlike the compression-ramp interactions discussed above, incident oblique shock interactions generally exhibit fluctuation amplification localised near the reflected shock foot. Accordingly, weakly separated cases show only a single local maximum \cite{Volpiani2018EffectsNonadiabatic},  whereas stronger interactions can develop two distinct peaks due to the combined effects of separation-shock motion and reattachment dynamics \cite{volpiani2020effects,dupont2006space,laguarda2024reynolds}. On the other hand, normalization with $\tilde{p}_w$ removes the bias introduced by the overall pressure rise and highlights the intrinsic unsteadiness relative to the local flow state. For the attached case R10 and the incipient separation case R20, a single peak is observed immediately downstream of the compression corner. In contrast, the fully separated cases R30 and R34 exhibit a distinct two-peak structure. The first, located slightly downstream of the separation point, reaches $30\%$ and $27.6\%$ of the local mean pressure for R30 and R34, respectively, and is associated with the low-frequency unsteadiness of the separation shock foot. The second, larger peak downstream of reattachment attains higher values of $34.5\%$ and $33.54\%$ of the local mean pressure, respectively, and is attributed to the unsteady shear-layer impingement.

The intensity of $p_{\text{rms}}/p_\infty$ is further compared against the theoretical relation proposed by Laganelli et al. \cite{laganelli1983wall} for wall-pressure fluctuations beneath compressible turbulent attached boundary layers, given by,
\begin{equation} \label{eq:eq_prms}
\frac{p_{\mathrm{rms}}}{q_\infty}
= \frac{0.006}
{\left[0.5 + ({T_w}/T_{r}) (0.5+0.09M_\infty^2) + 0.04M_\infty^2\right]^{0.64}},
\end{equation}

where $q_\infty=\rho_\infty u_\infty^2/2$ is the free-stream dynamic pressure and $T_r$ is the adiabatic recovery wall temperature. A range of numerator values in Eq. \ref{eq:eq_prms} have been suggested in the literature, including $0.009$ by Beresh et al. \cite{beresh2011fluctuating} and $0.008$ suggested by Ritos et al. \cite{ritos2019acoustic} based on the simulations for flows from Mach 4--8. Using the latter value, Ritos et al. \cite{ritos2020computational} reported encouraging agreement with their $p_{\mathrm{rms}}$ predictions. In the present case, however, the modified relation with a numerator value of 0.009 
remains in close agreement with the present LES predictions. \textcolor{black}{As shown by the black circles in Fig. \ref{fig:prms_plots} (a), the $p_{\text{rms}}$ level in the attached boundary layer is reproduced to within $\approx4\%$.}

\begin{figure}[!t]
    \centering
    \includegraphics{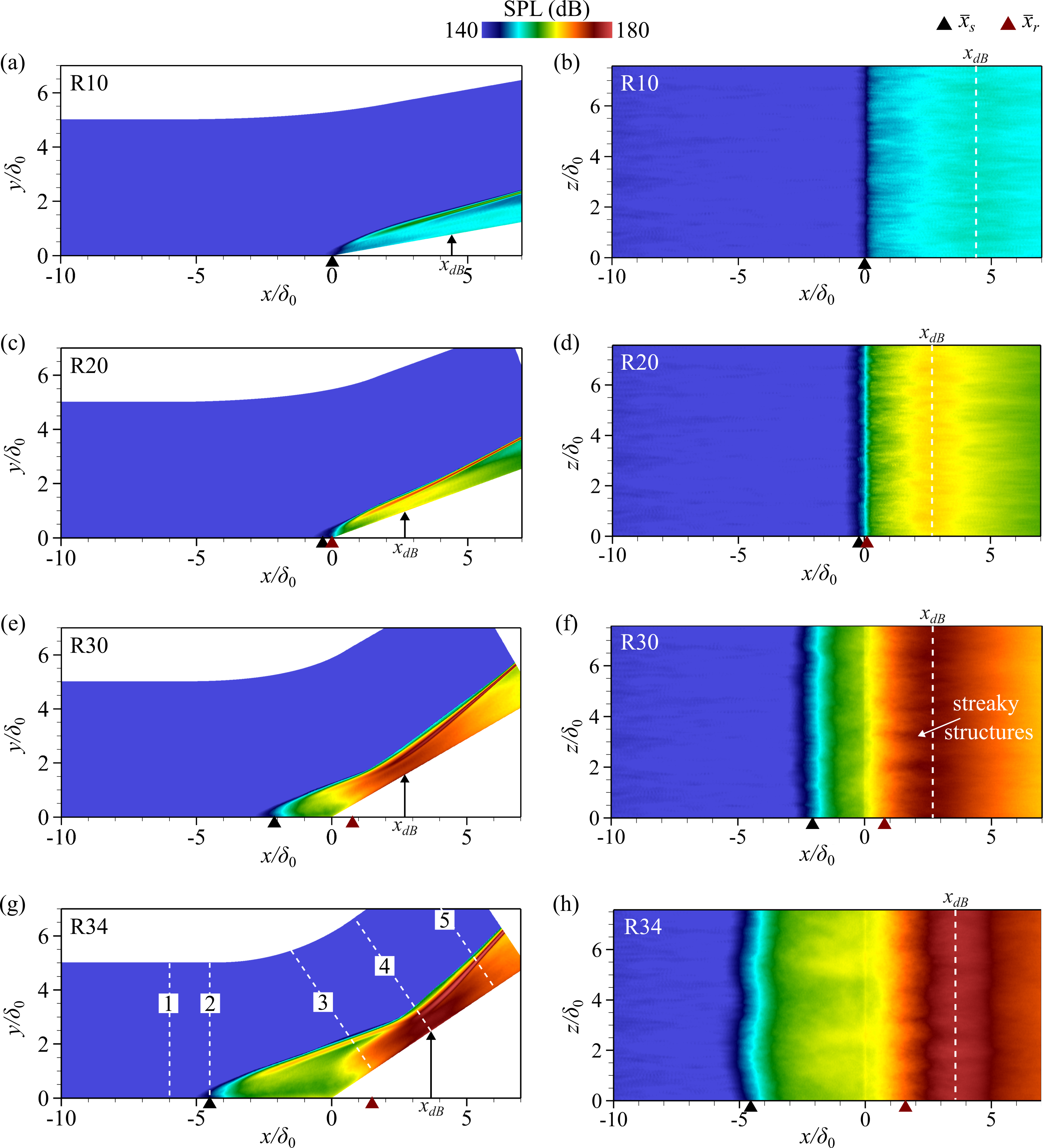}
    \caption{Contours of overall sound pressure level (OASPL) in decibels in the wall-normal ($xy$) and spanwise ($xz$) planes for different ramp angles, with the $xy$ plane shown on the left and the $xz$ plane on the right. Triangle symbols indicate mean separation (red) and reattachment (black) locations. The streamwise positions of the peak OASPL loads are indicated by $x_{dB}$.}
    \label{fig:OASPL_contour}
\end{figure}


\subsubsection{Quantification of Acoustic Loading}
To quantify the physical impact of these pressure fluctuations on the compression ramp, we next examine the associated acoustic loading. This is characterized using the overall sound pressure level (OASPL), defined in decibel scale and is calculated as $\text{OASPL} = 20\text{log}_{10}(p_{\mathrm{rms}}/p_{0})$ dB, where $p_0=20\mu \text{Pa}$ is the reference pressure. Figure \ref{fig:OASPL_contour} shows the contours of OASPL in decibels in the wall-normal ($xy$) and spanwise ($xz$) planes for different ramp angles, with the $xy$ plane shown on the left and the $xz$ plane on the right. The mean separation and reattachment locations are also shown. The streamwise positions of the peak OASPL loads are indicated by $x_{dB}$. In all cases, the acoustic levels remain relatively low and nearly uniform within the undisturbed incoming turbulent boundary layer, with amplification emerging near the separation location and increasing progressively through the interaction region. In the separated cases, the maximum OASPL consistently occurs downstream of reattachment, aligned with the peak location of $p_{rms}/p_{\infty}$. It is worth noting that the peak loading also aligns with the spatial location where the separation shock intersects with the reattachment shocks. In this region, the unsteadiness associated with shear-layer impingement and rapid recompression amplifies wall-pressure fluctuations. Further downstream, the acoustic levels decay gradually as the boundary layer begins to recover, reflecting the progressive depletion of energetic eddies in the post-reattachment region. For the attached case, the OASPL amplification remains confined to a narrow region near the corner, with minimal downstream decay. Additionally, all cases show higher OASPL values near the shock system. The wall ($x-z$) OASPL distributions exhibit a broadband spanwise character across all ramp configurations, with distinct streamwise bands of varying intensity distributed over the span. Within each band, localised regions of relatively high and low acoustic loading are observed. \textcolor{black}{Marginally elevated OASPL levels of $\approx$ 1 dB above the downstream levels ($x/\delta_0\lesssim -5$), are also evident.} This increase is attributed to the unsteadiness of the incoming oblique shock at the inflow plane. \textcolor{black}{A streak-like structure is also observed near reattachment and appears to originate from coherent vortical structures within the reattaching shear layer, which produce spanwise-irregular regions of high acoustic loading upon impingement on the wall.} Moreover, a pronounced spanwise modulation emerges near the separation region for the largely separated cases, aligned with low-frequency shock motion.

\begin{figure}[htpb]
    \centering
    \includegraphics{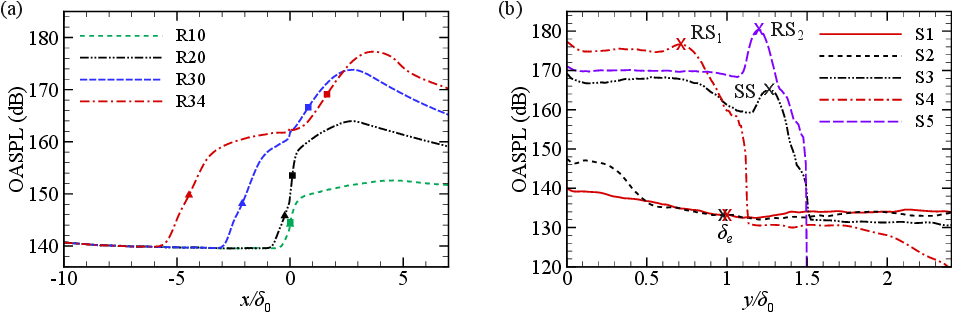}
    \caption{Profiles of (a) streamwise evolution of spanwise-averaged wall OASPL (dB) for different ramp angles, and (b) wall-normal distribution of OASPL (dB) for R34 case at different streamwise stations (1–5). Refer to the Fig. \ref{fig:OASPL_contour} for station nomenclature.}
    \label{fig:OASPL_plots}
\end{figure}

Figure \ref{fig:OASPL_plots} (a) presents the streamwise evolution of the spanwise-averaged wall OASPL for all four ramp configurations. Upstream of the interaction, the levels remain close to $140$ dB, in agreement with the values reported for an undisturbed turbulent boundary layer \cite{ritos2020computational}. Through the interaction region, the OASPL rises monotonically, reaching peak values of $152.6$, $164.0$, $173.8$, and $177.3$ dB for R10, R20, R30, and R34, respectively. For the separated cases, the location of maximum loading, $x_{dB}$, shifts progressively downstream, reaching approximately $2.7\delta_0$ for R30 and $3.8\delta_0$ for R34. The strongest interaction case (R34) exhibits progressive amplification, with OASPL levels rising by $9.8$ dB from the upstream level to the mean separation, $19.2$ dB from separation to reattachment, and a further $8.3$ dB to the global maximum. Downstream of the peak, the OASPL gradually decays for R20, R30, and R34, \textcolor{black}{whereas levels for R10 shows marginal decay, indicating faster recovery of the attached boundary layer within the computational domain. Given the limited availability of OASPL data for high-speed shock-induced interactions, the present dataset extends the existing reference base to strongly separated hypersonic compression-ramp flows across multiple interaction strengths. Nevertheless, the predictions of OASPL levels are broadly consistent with the limited literature on comparable high-speed interactions.} For instance, Ritos et al. \cite{ritos2020computational} reported upstream TBL levels of 140 dB increasing to more than 170 dB downstream of the reattachment region for a Mach $7.2$ expansion-compression ramp. Similarly, Zuo et al.~\cite{zuo2023wall} observed OASPL levels in the range of $\approx160-173$ dB in Mach 2.05 conical SWBLI.


Figure \ref{fig:OASPL_plots} (b) plots the wall-normal distribution of OASPL for the R34 case at five representative streamwise stations (indicated in Fig. \ref{fig:OASPL_contour}). Station 1 lies within the undisturbed upstream boundary layer, station 2 near the mean separation point, station 3 at reattachment, station 4 at the peak wall OASPL location $x_{dB}$, and station 5 is further downstream at $x/\delta_0 = 6$. At station 1, the OASPL decreases up to the boundary-layer edge $\delta_e$, followed by a slight increase in its vicinity and then settles into a near constant plateau. This marginal increase ($\approx0.6$ dB) is likely associated with the weak oblique shock introduced by the inflow condition. \textcolor{black}{In the present study, the local boundary-layer edge thickness, $\delta_e$, is calculated using a vorticity-magnitude-based threshold.} At the separation location (station 2), a slight near-wall reduction (0.95 dB) is followed by a local maximum, after which the OASPL decreases towards the outer region and recovers the same outer-layer plateau as observed at the upstream station. \textcolor{black}{At stations 3--5, a more pronounced decrease is observed, with acoustic levels reducing by 2.2 dB, 2.3 dB, and 1.2 dB, respectively, with increasing wall-normal distance. This is followed by an off-wall peak associated with the separation shock (SS) at station 3 and the reattachment shocks, RS1 and RS2, at stations 4 and 5, respectively.} Notably, while the off-wall maxima at stations 3 and 4 remain below the corresponding wall OASPL levels, station 5 exhibits the largest vertical peak of $182.7$ dB against a wall value of $170.95$ dB. This observation is in line with the increased pressure fluctuation intensity reported by Zuo et al. \cite{zuo2023wall} for shock-induced separation in a conical SWBLI at $M_\infty=2.05$. \textcolor{black}{Beyond this peak, a rapid decay in OASPL is observed. At stations 3 and 4, the profiles subsequently exhibit a region of nearly constant OASPL before decreasing toward the outer region.} 

\begin{figure}[htpb]
    \centering
    \includegraphics{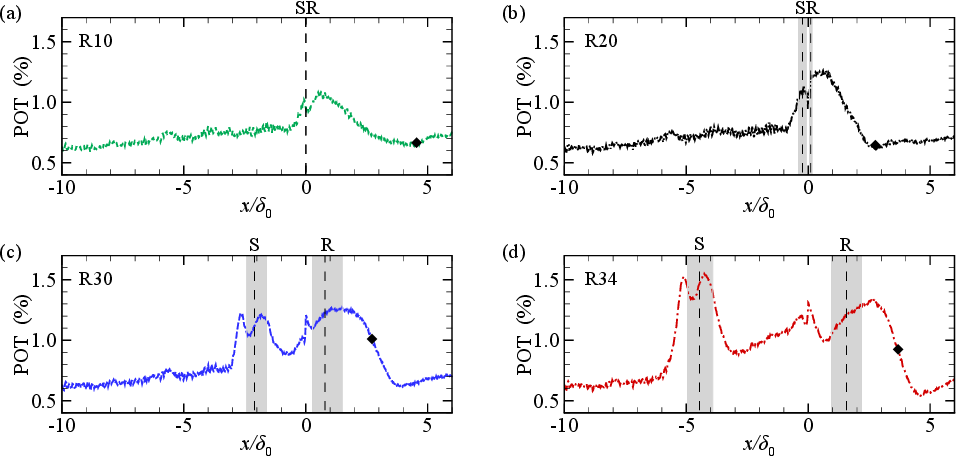}
    \caption{Streamwise distribution of peak-over-threshold (POT) exceedance of wall pressure fluctuations ($p'$) for (a) R10, (b) R20, (c) R30, and (d) R34. Grey bands indicate the temporal variation of spanwise-averaged separation and reattachment, with dashed lines marking the mean positions. Symbols denote peak acoustic loading location.}
    \label{fig:POT_plots}
\end{figure}

\subsubsection{Nature of Acoustic Loading}
\textcolor{black}{While the preceding analysis established the spatial distribution and intensity of acoustic loading, the nature of the loading environment, whether relatively uniform or dominated by intermittent in nature, remains to be characterised. This distinction has direct implications for acoustic fatigue assessment and structural design considerations. To characterise the loading environment, the temporal intermittency of the wall-pressure fluctuations is examined through a peak-over-threshold exceedance metric.} This metric is based on the intermittency factor introduced by Dolling and Murphy \cite{dolling1983unsteadiness}, which characterises the fraction of time the flow is disturbed by the shock. The following method uses a fixed upstream reference threshold to distinguish between undisturbed and shock-affected states. \textcolor{black}{In the present work, however, the interest is not in identifying shock-passage events but rather in quantifying the fraction of time the wall-pressure fluctuations exceed a threshold associated with burst-like extreme events.} To this end, a locally adaptive threshold based on the local $p_{rms}$ at each streamwise location is adopted. The exceedance (POT) is defined as
 
\begin{equation}
\text{POT\,}(x)=\frac{\mathrm{time}\left[p_w'(x) > np_{rms}(x)\right]}{\mathrm{total\ time}}\times 100,
\end{equation}
\textcolor{black}{where $n=3$ is the threshold parameter, corresponding to pressure events that exceed three standard deviations above the local $p_{rms}$ \cite{dolling1983unsteadiness}}. Figure \ref{fig:POT_plots} compares the streamwise distribution of the peak-over-threshold (POT) exceedance of $p_w'$ at different ramp angles. The temporal variation of spanwise-averaged separation and reattachment is indicated by grey bands, and dashed lines mark their mean positions. The location of the peak $p_{rms}$ for each case is indicated by a diamond symbol. In the attached and incipient-separation cases, the exceedance rises upstream of the corner as the boundary layer responds to the adverse pressure gradient, forming a first local maximum. This is followed by a decrease and then a second, marginally stronger peak of $\approx1.05\%$ and $\approx1.25\%$ for R10 and R20, respectively, located downstream of the compression corner, beyond which the exceedance decays to a near-uniform plateau. On the other hand, for largely separated cases, a multi-peak structure with four distinct local maxima emerges, spanning from the shock-intermittency region to the reattachment vicinity. Although R30 exhibits nearly comparable peaks near separation and reattachment ($\approx1.25\%$), R34 exhibits a clear dominance of the separation-region peak, which reaches $\approx1.55\%$. This trend indicates that the most intense intermittent pressure events increasingly align with the low-frequency oscillation of the separation shock foot. In contrast, the peak $p_{\mathrm{rms}}$ locations consistently register lower POT values of $\approx0.67\%$, $\approx0.65\%$, $\approx1.0\%$ and $\approx0.9\%$ for R10, R20, R30 and R34, respectively, demonstrating that the elevated fluctuation intensity downstream of reattachment is relatively sustained in nature rather than intermittent. This spatial offset between peak intermittency and peak $p_{rms}$ loading indicates that two distinct surface-loading mechanisms operate across the interaction zone, each posing a different structural risk. \textcolor{black}{Impulsive loading at separation, though intermittent, can generate high-amplitude events that excite structural modes. When the dominant low-frequency content aligns with the natural frequencies of compliant panels, resonance may occur, leading to fatigue or even failure \cite{mcnamara2011aeroelastic, kokkinakis2020direct, zhang2022direct}. In contrast, the reattachment region experiences the strongest acoustic loading. It is characterised by relatively sustained fluctuations that constitute a classical high-cycle fatigue environment, in which damage is governed by a large number of loading cycles over the vehicle operational lifetime \cite{zhou2017frequency}. The present results, therefore, suggest that RMS-based acoustic loading criteria applied uniformly across the interaction zone may not adequately capture the distinct fatigue mechanisms associated with separation and reattachment.}


\begin{figure}[!t]
    \centering
    \includegraphics{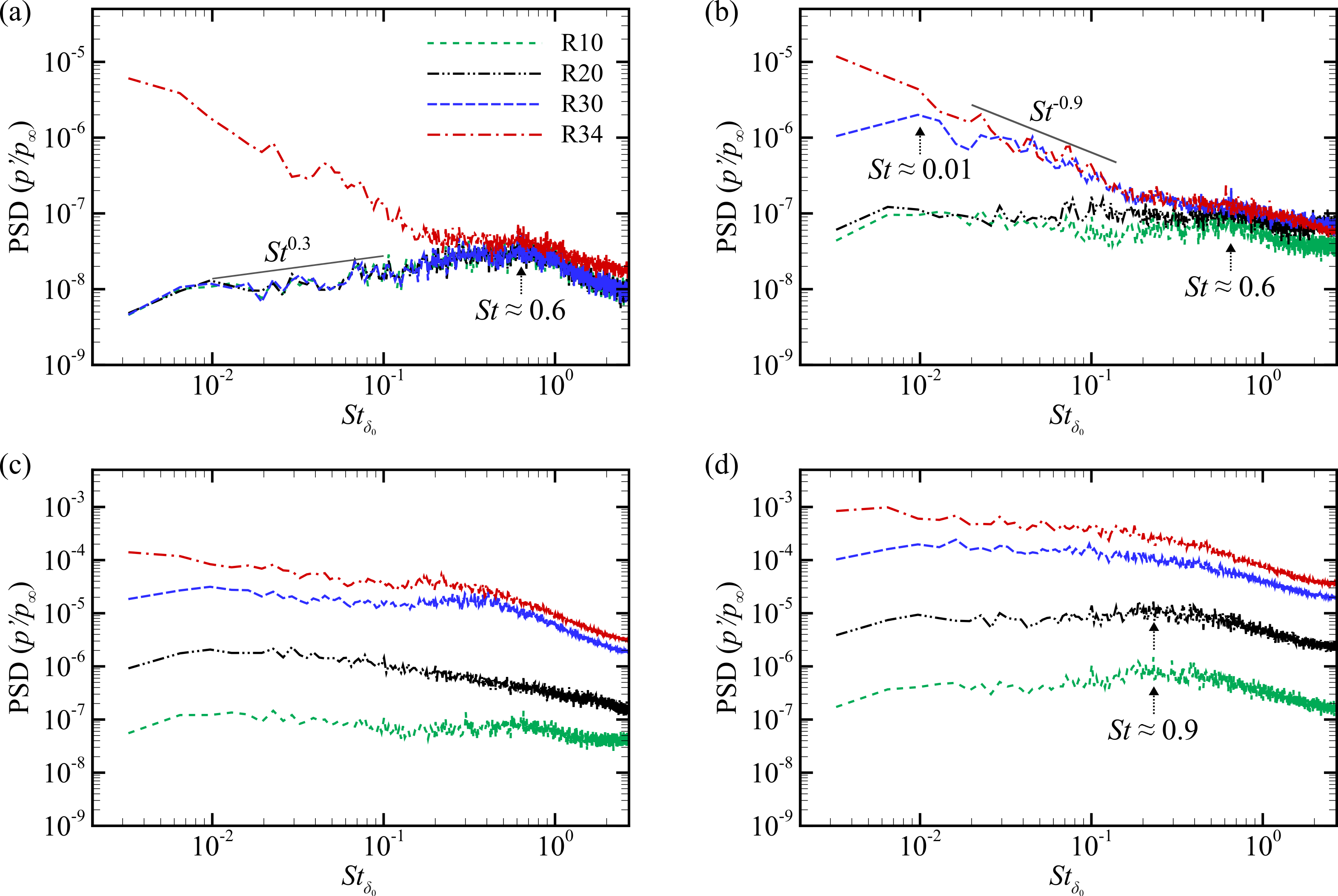}
    \caption{Spanwise averaged power spectral density (PSD) of pressure fluctuations for different ramp angles at selected streamwise locations of (a) $x/{\delta_0}=-5$, and the ramp angle specific locations corresponding to (b) separation point $x_s$, (c) reattachment point $x_r$, and (d) peak acoustic loading $x_{dB}$.}
    \label{fig:PSD_plots}
\end{figure}

\subsection{Spectral and Band-Resolved Characteristics}

With the acoustic loading footprint established, the spectral characteristics of the wall-pressure fluctuations are next examined using the power spectral density (PSD). A total of 4850 instantaneous two-dimensional wall-pressure slices are sampled at approximately $f_s\delta_0/u_\infty=5.5$. PSD is then estimated using Welch’s method with 3 segments and 50\% overlap using a Hamming window. \textcolor{black}{The resulting spectra are subsequently averaged in the spanwise direction to reduce the high-frequency oscillations and obtain a smoother spectral distribution.} Figure \ref{fig:PSD_plots} presents the power spectral density (PSD) of the normalized wall-pressure fluctuations, with frequency expressed as the Strouhal number $St =f\delta_0/u_\infty$. Spectra are shown for all ramp angles at representative streamwise locations corresponding to (a) the upstream undisturbed boundary layer at $x/\delta_0=-5$, (b) the separation point $x_s$, (c) the reattachment point $x_r$, and (d) the location of peak acoustic loading $x_{dB}$. 

At the upstream location (Fig.  \ref{fig:PSD_plots} a), the spectra for R10, R20 and R30 collapse closely and exhibit a characteristic peak near $St\approx0.6$, similar to that observed in canonical turbulent boundary-layer wall-pressure spectra reported in the literature \cite{zuo2023wall,kang2024direct,porter2019selective}. At lower frequencies, the undisturbed TBL spectrum follows $St^{0.3}$. This behaviour is broadly in agreement with the theoretical arguments of Ffowcs Williams \cite{williams1965surface} for compressible wall-pressure spectra in the $\omega\rightarrow0$ limit and also agrees with the $\omega^{0.3}$ scaling reported by Ritos et al. \cite{ritos2020computational} under similar flow conditions. In comparison, R34 shows noticeable low-frequency amplification, as the nominally upstream probe falls within the intermittent shock-influence region, where large-scale shock excursions start to modulate the wall-pressure signal. More pronounced differences emerge at the separation location (Fig. \ref{fig:PSD_plots}b). For R10 and R20, the low-frequency spectrum approaches a near-constant plateau ( $St^0$), indicating that the shock foot motion remains weak and is not reflected strongly on the wall-pressure signature. In contrast, R30 and R34 exhibit strong low-frequency amplification with an $\approx St^{-0.9}$ behaviour. In particular, R30 shows a distinct peak near $St\approx0.01$, corresponding to characteristic separation-shock motion frequencies reported for compression-ramp interactions \cite{Zhang2024EffectsWallTemperature}. At reattachment (Fig. \ref{fig:PSD_plots}c), the spectral amplitude increases across nearly the entire frequency range with increasing ramp angle, indicating broadband amplification associated with stronger shear-layer impingement and recompression \cite{dupont2006space}. A similar broadband increase is also observed at the peak acoustic-loading location (Fig. \ref{fig:PSD_plots}d). Notably, R10 and R20 retain a weak spectral peak near $St\approx0.9$, whereas the more strongly separated R30 and R34 cases exhibit a monotonic decay, reflecting a shift toward broadband spectral character as the interaction strength increases. 



\begin{figure}[!t]
    \centering
    \includegraphics{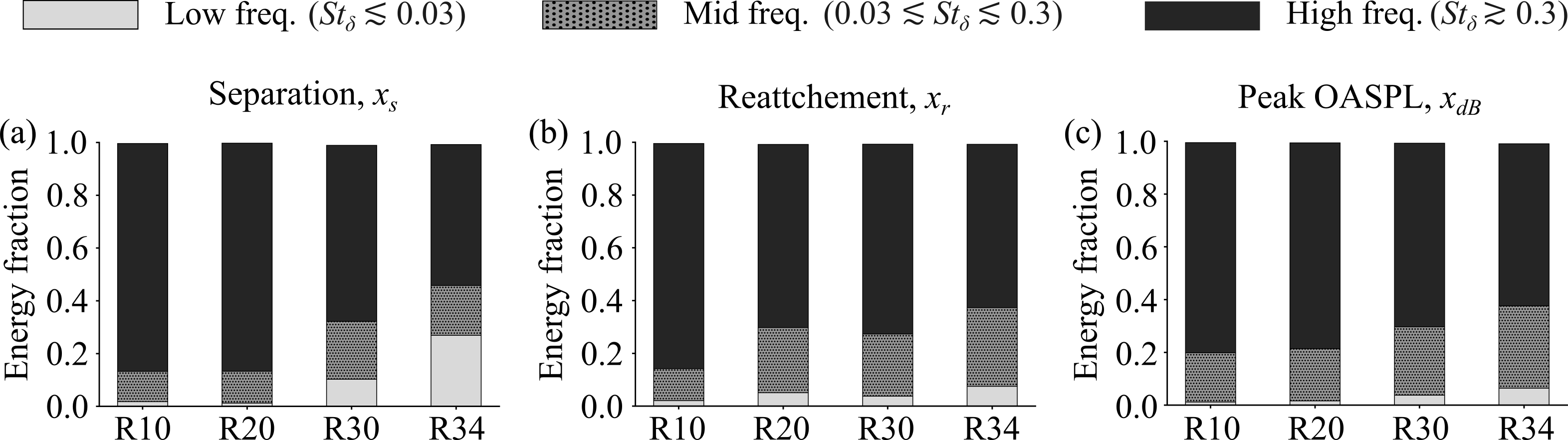}
    \caption{Fractional spectral energy contributions from the band-pass filtered low-, mid-, and high frequency signals at the ramp angle specific locations corresponding to (a) separation point $x_s$, (b) reattachment point $x_r$, and (c) peak acoustic loading.}
    \label{fig:PSD_contri}
\end{figure}

To quantify the spectral redistribution observed in the PSD, the fluctuating wall-pressure signal $p'$ is decomposed into three frequency bands using band-pass filtering: low-frequency ($St_{\delta}\lesssim0.03$), mid-frequency ($0.03\lesssim St_{\delta}\lesssim0.3$), and high-frequency ($0.3\lesssim St_{\delta}\lesssim2.7$). These bands correspond to interaction breathing, coherent shear-layer fluctuations, and integral-scale turbulence, respectively, following the frequency ranges identified in previous SWBLI studies \cite{adler2020dynamics, adler2022influence, gaitonde2023dynamics, suri2024one}. The upper cutoff of $St_{\delta}=2.7$ represents the resolved frequency limit of the present simulation. Figure \ref{fig:PSD_contri} presents the resulting fractional band energy at the (a) separation location $x_s$, (b) reattachment location $x_r$, and (c) point of peak acoustic loading $x_{dB}$. At separation (Fig. \ref{fig:PSD_contri}a), the high frequency band remains dominant across all cases, although its contribution decreases progressively with increasing ramp angle, accompanied by increases in the low and mid frequency bands. \textcolor{black}{The most noticeable change occurs in the low frequency band, where the energy fraction increases from 0.10 for R30 to 0.27 for R34, indicating the onset of large-scale shock foot unsteadiness and separation bubble breathing (identified in Fig. \ref{fig:PSD_plots}b).} This confirms that increasing interaction strength shifts spectral energy away from the canonical turbulent boundary-layer content toward low-frequency shock motion. At reattachment and peak acoustic loading (Fig. \ref{fig:PSD_contri},(b,c)), the high-frequency band remains the dominant contributor across all ramp angles. Its contribution, however, decreases with increasing ramp angle, accompanied by a corresponding rise in the mid-frequency band, while the low-frequency contribution remains comparatively smaller. For the weaker interactions (R10 and R20), the downstream spectral distribution is consistent with a recovering turbulent boundary layer, suggesting that the flow is relaxing towards a more canonical wall-turbulence state. 

The frequency-dependent energy distribution identified above has direct implications for existing wall-pressure spectral models. To the authors’ knowledge, the wall-pressure spectral model proposed by Laganelli et al. \cite{laganelli1993prediction} remains the only available predictive method for high-speed flows with shock-induced separation. It augments the undisturbed approach-flow PSD using a factor based on the inviscid shock strength to predict the spectra across the interaction region. While this formulation captures the overall increase in spectral energy, it assumes a spatially uniform amplification and does not distinguish between the separation and reattachment zones. The band-isolated results show that this assumption does not adequately capture the underlying spectral behaviour. The present dataset, spanning from attached to strongly separated interactions, provides a basis for extending such models to incorporate zone-dependent spectral behaviour or at least to capture the spatial evolution of the PSD in high-speed flows with pressure gradients.


\subsubsection{Band-isolated OASPL maps}

\begin{figure}[!t]
    \centering
    \includegraphics{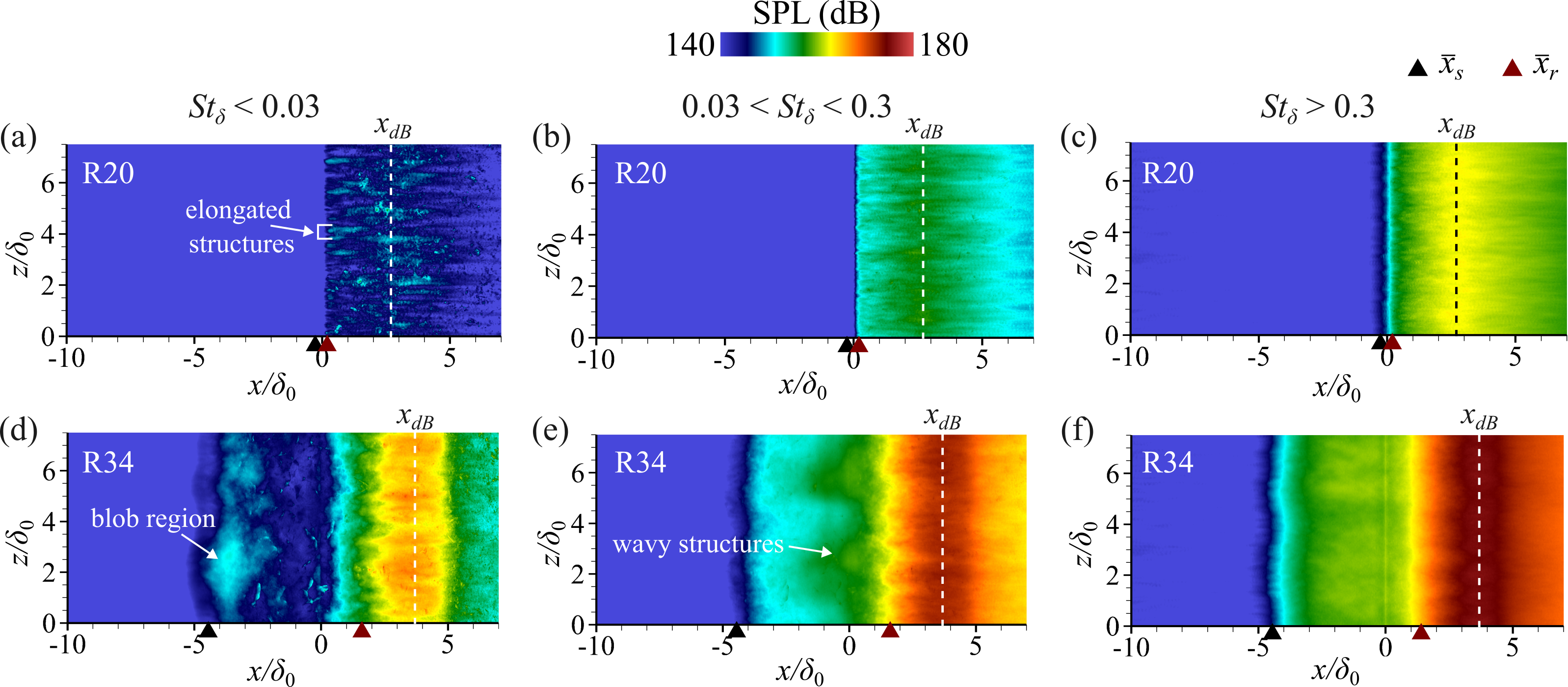}
    \caption{Contours of band-pass filtered low-, mid-, and high frequency overall sound pressure level (OASPL) in decibels on the $xz$ wall plane for R20 and R34 case. Triangle symbols indicate mean separation (red) and reattachment (black) locations. The streamwise positions of the peak OASPL loads are indicated by $x_{dB}$.}
    \label{fig:OASPL_bandp}
\end{figure}

Having established the relative spectral energy contributions at specific streamwise locations, the spatial distribution of each frequency band is examined through band-isolated overall sound pressure level (OASPL) maps. For each band, the OASPL is obtained by integrating the PSD of the fluctuating wall-pressure signal $p'$ over the corresponding frequency limits, rather than over the full resolved spectrum. This yields band-isolated OASPL distributions that highlight the wall-loading footprint of the low-, mid-, and high-frequency motions. Figure \ref{fig:OASPL_bandp} shows the contours of the band-pass-filtered low-, mid-, and high-frequency OASPL on the $xz$ wall plane for the R20 and R34 cases. These two configurations are selected to contrast the wall-pressure loading characteristics between a relatively weak interaction and a strongly separated interaction. Triangle symbols denote the mean separation (red) and reattachment (black) locations, while the streamwise positions of the peak band-isolated OASPL are marked by $x_{dB}$ and closely follow the trends observed in Fig. \ref{fig:OASPL_contour}. The band-specific observations are discussed below.

Low-frequency band: The low-frequency acoustic loading remains weak across the interaction domain for the R20 case, consistent with its incipient separation and the inability of the small recirculation bubble to sustain coherent low-frequency dynamics. Downstream of the compression ramp, streamwise-elongated streaks are observed, alternating in the spanwise direction. These likely represent the large-scale structures and motions convecting through the weak interaction. The R34 case, however, exhibits a pronounced blob-like region of elevated OASPL near separation, providing a clear acoustic signature of bubble breathing and large-scale shock oscillation. Moving downstream along the separation bubble, the low-frequency OASPL decreases progressively toward the compression corner. This weakening footprint further suggests a gradual shift from low-frequency energy towards broadband shear-layer dynamics \cite{laguarda2024reynolds}.

Mid-frequency band: For the R20 case, the mid-frequency acoustic loading closely resembles that of the low-frequency band, with streamwise-elongated streak-like structures. For R34, the OASPL levels in this band increase rapidly in both amplitude and spatial extent downstream, indicating substantially stronger mid-frequency acoustic radiation than in R20. The wavy structures observed within the separated region are aligned with the large-scale shear-layer instabilities prior to reattachment, whereas the intense loading downstream is associated with shear-layer impingement.

High-Frequency band: In both cases, the high-frequency band exhibits the highest OASPL levels among all and increases monotonically downstream. \textcolor{black}{However, in the R34 case, the OASPL associated with the low frequency band increases locally near separation onset and becomes comparable to that of the high frequency band, indicating enhanced low frequency acoustic energy in this region.}


\subsubsection{Band-isolated spatio-temporal maps}

\begin{figure}[t]
    \centering
    \includegraphics{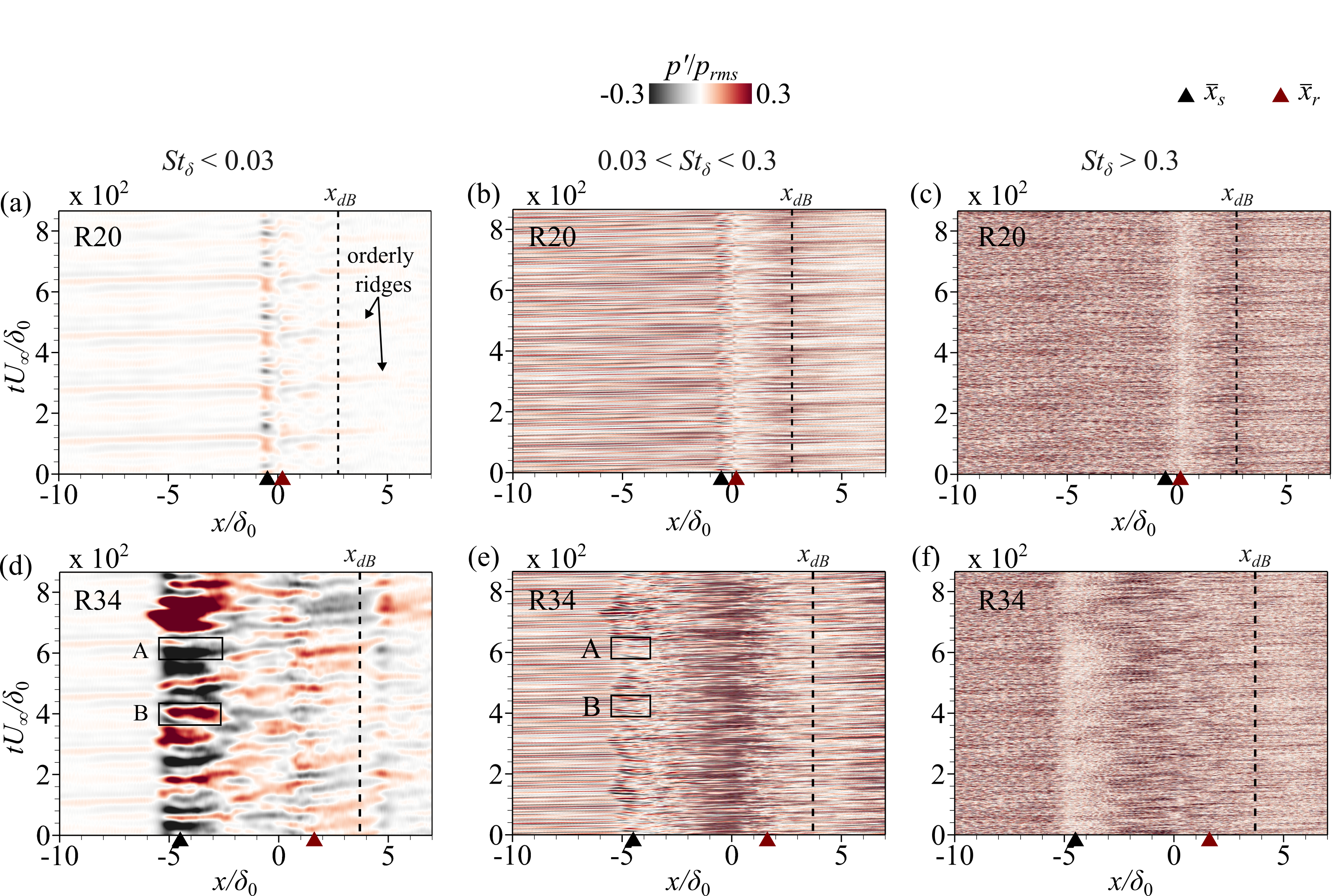}
    \caption{Spatio-temporal contours of band-pass filtered low-, mid-, and high frequency pressure fluctuations ($p'$) for R20 and R34 case. Triangle symbols indicate mean separation (red) and reattachment (black) locations. The streamwise positions of the peak OASPL loads are indicated by $x_{dB}$.}
    \label{fig:space_time_bandp}
\end{figure}

To determine whether the energetic structures identified in the band-isolated OASPL maps convect as organised structures, we next consider the corresponding spatio-temporal correlations of band-pass-filtered pressure fluctuations. \textcolor{black}{Similar analyses have previously been employed to identify coherent structures, and their spatio-temporal evolution \cite{musta2024investigation, adler2020dynamics}.} Figure \ref{fig:space_time_bandp} shows the spatio-temporal contours of band-pass filtered low-, mid-, and high frequency pressure fluctuations ($p'$) for R20 and R34 case. In the present study, the pressure fluctuations ($p'$) are normalized by the local $p_{rms}$ to \textcolor{black}{facilitate comparison} across the interaction zone, ensuring that the space-time correlations exhibit structural coherence rather than local energy content. The band-specific observations are discussed below.

Low-frequency band: Both R20 and R34 exhibit elongated alternating streaks of positive and negative pressure fluctuations that originate within the upstream turbulent boundary layer and convect downstream across the interaction onto the compression ramp. Several of these structures remain continuous from upstream and are further amplified through the interaction. \textcolor{black}{A notable feature is the local phase reversal across the interaction, whereby positive fluctuations in the intermittent region are often followed by negative fluctuations over the ramp surface, and vice versa. For instance, the black dashed region labelled as A and B corresponds to negative (black) and positive (red) pressure fluctuations (Fig. \ref{fig:space_time_bandp}d).  Downstream of these regions, the pressure fluctuations are predominantly associated with fluctuations of opposite sign on the compression ramp surface ($x/\delta_0 > 0$) \cite{musta2024investigation}.} Compared with the R20 case, R34 displays distinct behaviour near separation, consisting of large, temporally extended structures of alternating positive and negative pressure that appear nearly vertical in the spatio-temporal plane, indicating an oscillatory rather than convective nature. This behaviour is a clear signature of separation-bubble breathing and low-frequency shock motion, with the shock foot periodically moving upstream and downstream and extending over a wider range of timescales. Downstream of reattachment, both R20 and R34 exhibit inclined structures consistent with convective behaviour. In the former case, the shorter separation bubble confines the low-frequency shock-foot motion to a narrow region. As a result, the structures convect from separation to reattachment over a short streamwise distance, with minimal distortion. This produces narrow, orderly ridges in the space–time contour. In the latter case, the longer separation bubble promotes sustained interaction between the shear layer and the unsteady shock system. This leads to a loss of coherence near reattachment, producing broader and disordered ridges.


\begin{figure}[htpb]
    \centering
    \includegraphics{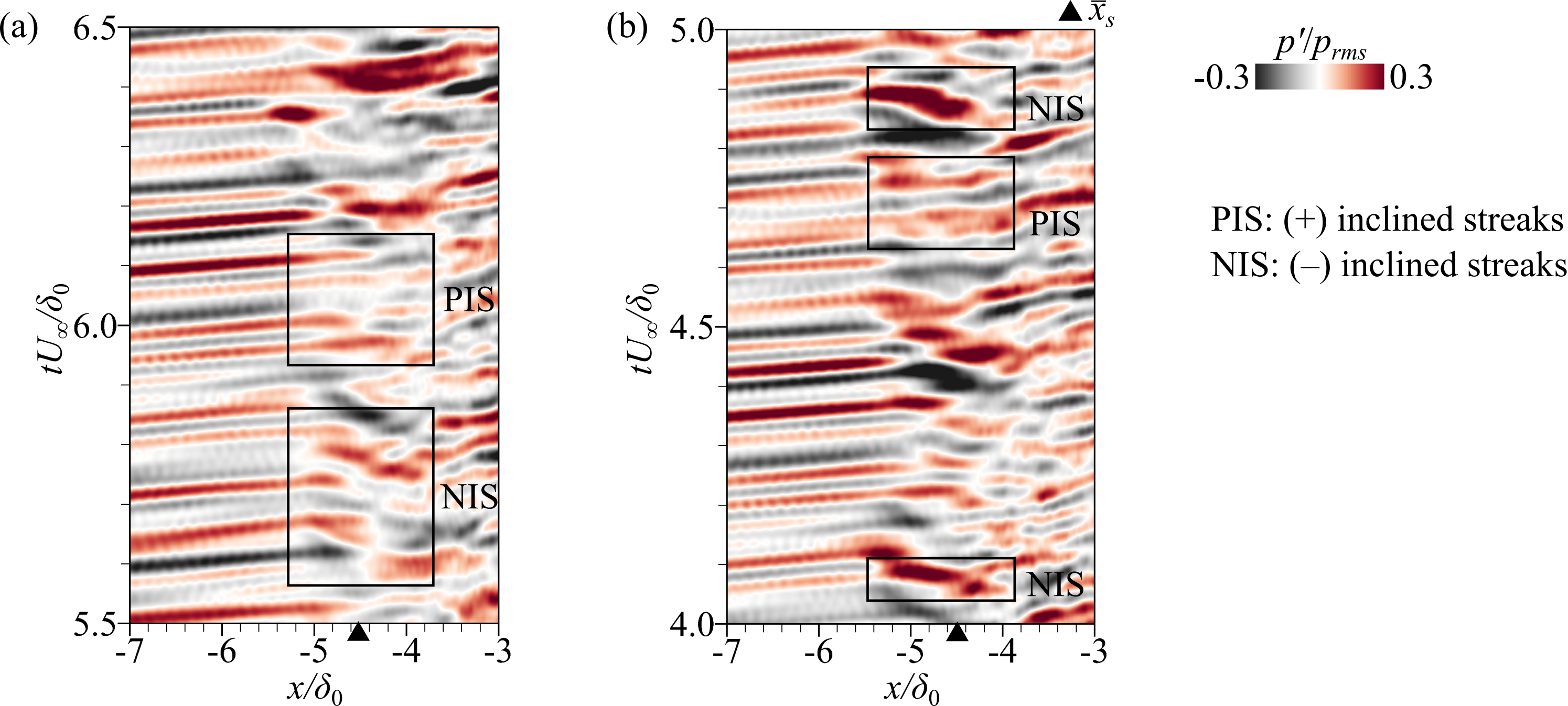}
    \caption{Magnified spatio-temporal contours of band-pass filtered mid-frequency pressure fluctuations ($p'$) for the R34 case, highlighting (a) region A and (b) region B (indicated in Fig. \ref{fig:space_time_bandp}e).}
    \label{fig:34d_mid_zoom}
\end{figure}

Mid-frequency band: The fluctuation field in this band is dominated by elongated, streaky structures that convect downstream and amplify through interactions. For R20, these structures remain almost continuous through the interaction, with marginal amplification observed over the incipient separation bubble. This modest growth suggests that the small recirculation bubble primarily acts as a weak amplifier of incoming disturbances, rather than a genuine source of coherent mid-frequency dynamics. For R34, the response is markedly richer, with a clear signature of the separation shock evident near the onset of interaction. Three distinct behaviours are observed in this region: amplified stationary structures, positively inclined structures, and negatively inclined structures (magnified view shown in Fig. \ref{fig:34d_mid_zoom}). The stationary streaks reflect shock-foot oscillations, indicating that the shock motion possesses sufficient broadband content to extend beyond the low-frequency band. The positively inclined streaks (Fig. \ref{fig:34d_mid_zoom}) are convective in character, representing Kelvin–Helmholtz vortices arising from the inflectional instability of the separated shear layer and propagating downstream from the separation region. Conversely, negatively inclined streaks (Fig. \ref{fig:34d_mid_zoom}) observed in the vicinity of the shock foot appear intermittently and indicate upstream-propagating pressure disturbances generated by the interaction between downstream convecting Kelvin–Helmholtz vortices and the shock foot. \textcolor{black}{A similar bidirectional (upstream and downstream) propagation of disturbances was reported by Pirozzoli and Grasso \cite{pirozzoli2006direct} in DNS of an impinging SWBLI at Mach 2.25. The upstream-propagating component was interpreted as an acoustic feedback process generated by shock-vortex interaction near the impinging shock foot and is linked to the observed low-frequency unsteadiness. The coexistence of oppositely directed $p'$ streaks near separation in the present spatio-temporal map suggests that a similar feedback mechanism may be active in the interaction.}

High-frequency band: In the high-frequency band, R20 shows a marginal attenuation of fluctuation amplitude at the compression corner, followed by downstream regrowth toward the peak acoustic loading region and subsequent decay. In the R34 case, the amplitude initially weakens near the onset of the interaction region, where the dynamics are dominated by low-frequency motion. Interestingly, the most intense high-frequency streaks are concentrated within the reverse-flow region, upstream of the compression corner ($-3\lesssim x /\delta_0 \lesssim 0$), likely reflecting leakage of mid-frequency content into the high-frequency band. The shock footprint in this band remains relatively weak, and unlike in the R20 case, no distinct amplification is observed near the peak acoustic loading. 

\section{Conclusion} \label{conclusion}

In this study, we have investigated the effect of wall pressure fluctuations and the associated acoustic loading beneath a hypersonic boundary layer approaching a compression corner. To this end, an in-house high-order solver, COMPSQUARE, is employed to perform high-fidelity large-eddy simulations. The computations are carried out at a free-stream Mach number $M_{\infty}=6.04$ and an inlet momentum-thickness Reynolds number of $Re_{\theta}=4343$. The compression-corner angles are varied over $10^\circ$, $20^\circ$, $30^\circ$, and $34^\circ$ to assess the role of interaction strength. The flow conditions are in accordance with the experimental setup of Whalen et al. \cite{whalen2019unsteady}. Henceforth, the state of the upstream turbulent boundary layer is examined and found to agree favourably with numerical reference data and established correlations. The effect of compression corner angles on the resulting SWBLI characteristics is further explored. With increasing interaction strength, the flow transitions from an attached state (R10) to incipient separation (R20) and eventually to fully separated states (R30 and R34). \textcolor{black}{The attached and incipient separation cases show favourable agreement with the experiments \cite{whalen2019unsteady}, while the largely separated case exhibits a modest overprediction of the separation bubble.} 

The fluctuating wall-pressure intensity increases markedly with interaction strength, with the peak $p_{rms}$ rising by $\approx312\%$ from R20 to R30 and by a further $\approx67\%$ from R30 to R34. For the separated cases, this peak consistently occurs downstream of reattachment. The physical impact of pressure fluctuations is further quantified through overall sound pressure level (OASPL) maps. The wall OASPL distributions exhibit a broadband spanwise character across all ramp configurations, with streamwise bands of varying intensity. Within each band, localised regions of high and low acoustic loading are observed. In the strongest interaction (R34), the OASPL increases from 140 dB in the upstream turbulent boundary layer to a peak of 177.3 dB downstream of reattachment. Point over threshold (POT) statistics reveal that intense intermittent pressure events increasingly align with the low-frequency oscillation of the separation shock foot. In contrast, the peak fluctuation intensity downstream of reattachment is relatively sustained rather than intermittent, indicating a high-cycle fatigue loading environment. These findings suggest that uniformly applied RMS-based acoustic loading criteria may not adequately represent the distinct fatigue mechanisms governing separation and reattachment.
 
The temporal spectra of wall pressure fluctuations reveal a shift from turbulence-dominated (R10, R20) high frequencies to broadband low-frequency motion (R30, R34) with increasing interaction strength. The wall-pressure fluctuation signal is decomposed into low-, mid-, and high-frequency components based on the spectral regimes identified in the PSD analysis. Subsequently, band-isolated OASPL maps are computed to examine the spatial distribution of acoustic loading within each frequency band. The OASPL maps reveal high-frequency fluctuations as the dominant contributor across the interaction zone. \textcolor{black}{However, for the largely separated case of R34, low-frequency OASPL increases locally near separation onset and reaches levels comparable to those in the high-frequency band, indicating enhanced low-frequency acoustic loading in this region.} The low-frequency spatio-temporal map of $p'$ shows a signature of separation-bubble breathing and low-frequency shock motion, with the shock foot periodically moving upstream and downstream and extending over a wider range of timescales. By contrast, the mid-frequency band reveals downstream convecting Kelvin–Helmholtz structures and intermittent upstream propagating pressure waves near the shock foot. These upstream propagating waves are generated by shock–vortex interactions at the separation foot and propagate upstream within the interaction zone, further promoting shear-layer instability.





\section*{Acknowledgments}
The authors thank the National PARAM Supercomputing Facility (NPSF) for providing computing resources in the PARAM Siddhi cluster under the National Supercomputing Mission. We also acknowledge NSM for providing access to the computing resources of ‘PARAMRUDRA’ at PG Senapathy Centre, PlayField Ave, Indian Institute of Technology, Chennai, Tamil Nadu 600036, which is implemented by C-DAC and supported by the Ministry of Electronics and Information Technology (MeitY), and Department of Science and Technology (DST), Government of India. The authors would also like to acknowledge the support from the Prime Minister Research Fellowship (PMRF), Government of India. 

\bibliography{sample}

\end{document}